\documentclass[preprint,noshowpacs,noshowkeys,aps]{revtex4}%
\usepackage{amsfonts}
\usepackage{amsmath}
\usepackage{amssymb}
\usepackage{graphicx}%
\providecommand{\U}[1]{\protect\rule{.1in}{.1in}}
\begin{document}
\preprint{ }
\title[Nonlocal Phases of Local Wavefunctions]{Nonlocal Phases of Local Quantum Mechanical Wavefunctions in Static and
Time-Dependent Aharonov-Bohm Experiments}
\author{Konstantinos Moulopoulos}
\email{cos@ucy.ac.cy}
\affiliation{University of Cyprus, Department of Physics, 1678 Nicosia, Cyprus}
\keywords{Gauge Transformations, Aharonov-Bohm Effect}
\pacs{03.65.-w, 03.65.Vf, 03.65.Ta, 03.50.De}

\begin{abstract}
We show that the standard Dirac phase factor is not the only solution of the
gauge transformation equations. The full form of a general gauge function
(that connects systems that move in different sets of scalar and vector
potentials), apart from Dirac phases also contains terms of classical fields
that act nonlocally (in spacetime) on the local solutions of the
time-dependent Schr\"{o}dinger equation: the phases of wavefunctions in the
Schr\"{o}dinger picture are affected nonlocally by spatially and temporally
remote magnetic and electric fields, in ways that are fully explored. These
contributions go beyond the usual Aharonov-Bohm effects (magnetic or
electric). (i) Application to cases of particles passing through static
magnetic or electric fields leads to cancellations of Aharonov-Bohm phases at
the observation point; these are linked to behaviors at the semiclassical
level (to the old Werner \& Brill experimental observations, or their
"electric analogs" - or to recent reports of Batelaan \& Tonomura) but are
shown to be far more general (true not only for narrow wavepackets but also
for completely delocalized quantum states). By using these cancellations,
certain previously unnoticed sign-errors in the literature are corrected. (ii)
Application to time-dependent situations provides a remedy for erroneous
results in the literature (on improper uses of Dirac phase factors) and leads
to phases that contain an Aharonov-Bohm part and a field-nonlocal part: their
competition is shown to recover Relativistic Causality in earlier "paradoxes"
(such as the van Kampen thought-experiment), while a more general
consideration indicates that the temporal nonlocalities found here demonstrate
in part a causal propagation of phases of quantum mechanical wavefunctions in
the Schr\"{o}dinger picture. This may open a direct way to address
time-dependent double-slit experiments and the associated causal issues.

\end{abstract}
\volumeyear{2009}
\volumenumber{number}
\issuenumber{number}
\eid{identifier}
\date[June 14, 2010]{}
\maketitle

\section{\bigskip Introduction}

It is well established from Weyl's work (1929), but also from independent
proposals by Schr\"{o}dinger (1922), Fock (1927) and London (1927)\cite{Weyl},
that there exists a simple unitary (U(1)) phase mapping that connects
different quantum systems, when these are gauge-equivalent (and then the phase
that connects their wavefunctions is basically the gauge function of an
ordinary gauge transformation). A simple unitary mapping of this type is also
reserved for quantum systems moving in multiple-connected spacetimes (with
enclosed appropriately defined \textquotedblleft fluxes\textquotedblright\ in
the physically inaccessible regions) the corresponding \textquotedblleft gauge
transformation\textquotedblright\ termed singular, and the corresponding
\textquotedblleft gauge function\textquotedblright\ now being multiple-valued
(although the wavefunctions of the \textquotedblleft final\textquotedblright%
\ (mapped) system are still single-valued) leading to phenomena of the
Aharonov-Bohm type. In this paper we report on a phase mapping connecting
systems that are \textit{not} \textquotedblleft equivalent\textquotedblright%
\ (in the sense of the above two), since they can go through
\textit{different} classical fields in remote regions of space and/or time,
and we give explicit forms of the appropriate \textquotedblleft gauge
functions\textquotedblright. The results are exact, in analytical form, and
they generalise the standard Dirac phase factors derived from path integral
treatments (that are very often used in an incorrect way as we will
demonstrate); apart from a discussion of such misconceptions propagating in
the literature, we also give first actual applications of the new results in
static and time-dependent experiments, both of the Aharonov-Bohm type (i.e.
with inaccessible fields and their fluxes) but also with the particles
actually passing through classical fields, and even being in completely
general quantum states (and not necessarily narrow wavepackets in
semiclassical motion).

\section{\bigskip Motivation}

In order to motivate this paper let us first remind the reader of the standard
U(1) mapping%

\begin{equation}
\Psi_{2}(\mathbf{r},t)=e^{i\frac{q}{\hbar c}\Lambda(\mathbf{r},t)}\Psi
_{1}(\mathbf{r},t) \label{UsualPsi}%
\end{equation}
\ between the solutions of the time-dependent Schr\"{o}dinger (or Dirac)
equation for a quantum particle of charge $q$ that moves (as a test particle)
in two distinct sets of (predetermined and classical) vector and scalar
potentials that are connected with each other (through a gauge transformation)
via the \textquotedblleft gauge function\textquotedblright\ $\Lambda
(\mathbf{r},t)$, namely%

\begin{equation}
\nabla\Lambda(\mathbf{r,t})=\mathbf{A}_{2}(\mathbf{r},t)-\mathbf{A}%
_{1}(\mathbf{r},t)\qquad and\qquad-\frac{1}{c}\frac{\partial\Lambda
(\mathbf{r},t)}{\partial t}=\phi_{2}\left(  \mathbf{r},t\right)  -\phi
_{1}(\mathbf{r},t). \label{gaugetransf}%
\end{equation}
\ \qquad In the static case, and if for simplicity we start from system 1
being completely free of potentials ($\mathbf{A}_{1}=\phi_{1}=0$), the
wavefunctions of the particle in system 2 (moving in a vector potential
$A(\mathbf{r})$) will acquire an extra phase with an appropriate
\textquotedblleft gauge function\textquotedblright\ $\Lambda(\mathbf{r})$ that
must satisfy$\ $%

\begin{equation}
\nabla\Lambda(\mathbf{r})=\mathbf{A}(\mathbf{r}). \label{usualgrad}%
\end{equation}
\ 

The standard (and widely-used) solution of this is the line integral%

\begin{equation}
\Lambda(\mathbf{r})=\Lambda(\mathbf{r}_{\mathbf{0}})+\int_{\mathbf{r}_{0}%
}^{\mathbf{r}}\mathbf{A}(\mathbf{r}^{\prime})\boldsymbol{.}d\mathbf{r}%
^{\prime} \label{usualABLambda}%
\end{equation}
(which, by considering two paths encircling an enclosed inaccessible magnetic
flux, leads to the well-known magnetic Aharonov-Bohm effect\cite{AB}). It
should however be stressed that the above is only true if (\ref{usualgrad}) is
valid for \textbf{all} points $\mathbf{r}$ of the region where the particle
moves, i.e. if the particle in system 2 moves (as a narrow wavepacket) always
outside magnetic fields ($\nabla\times\mathbf{A}=0$ everywhere). Similarly, if
the particle in system 2 moves in a spatially homogeneous scalar potential
$\phi(t)$, the appropriate $\Lambda$ must satisfy\ \ \ \
\begin{equation}
-\frac{1}{c}\frac{\partial\Lambda(t)}{\partial t}=\phi(t), \label{usualdt}%
\end{equation}
the standard solution being%

\begin{equation}
\Lambda(t)=\Lambda(t_{0})-c\int_{t_{0}}^{t}\phi(t^{\prime})dt^{\prime}
\label{usualelectricABLambda}%
\end{equation}
that gives the extra phase acquired by system 2 (this result leading to the
electric Aharonov-Bohm effect\cite{AB} by applying it to two equipotential
regions, such as two metallic cages held in distinct time-dependent scalar
potentials). Once again, it should be stressed that the above is only true if
(\ref{usualdt}) \ (and the assumed spatial homogeneity of the scalar potential
$\phi$ and of $\Lambda$) is valid at \textbf{all} times $t$ of interest, i.e.
if the particle in system 2 moves (as a narrow wavepacket) always outside
electric fields ($\mathbf{E}=-\nabla\phi-\frac{1}{c}\frac{\partial\mathbf{A}%
}{\partial t}=0$ \ at all times). (In the electric Aharonov-Bohm setup, the
above is ensured by the fact that $t$ lies in an interval of a finite duration
$T$ \ for which the potentials are turned on, in combination with the
narrowness of the wavepacket$\boldsymbol{;}$ this guarantees that, during $T$,
the particle has vanishing probability of being at the edges of the cage where
the potential starts having a spatial dependence. The reader is referred to
Appendix B of Peshkin\cite{Peshkin} that demonstrates the intricasies of the
electric Aharonov-Bohm effect, to which we return with an important comment at
the end of the paper (Section XI)).

\bigskip

In the present work, we relax the above assumptions and present more general
solutions of the system of Partial Differential Equations (\ref{gaugetransf}),
covering cases where the particle is \textit{not} necessarily a narrow
wavepacket (it can actually be in completely delocalized states) and is
\textit{not} excluded from \textit{remote} regions (in space-time) of
nonvanishing (or, more generally, of unequal) fields (magnetic or electric),
regions therefore that are actually accessible to the particle (hence
non-Aharonov-Bohm cases, or even combinations of spatial multiple-connectivity
of the magnetic Aharonov-Bohm type, but simultaneous simple-connectivity in
spacetime (i.e. in the $(x,t)-$plane)). We find analytically nonlocal
influences of these remote fields on $\Lambda(\mathbf{r},t)$ (with
$(\mathbf{r},t)$ the observation point in spacetime), and therefore on the
phases of wavefunctions at $(\mathbf{r},t)$, that seem to have a number of
important consequences$\boldsymbol{:}$ they provide (i) a natural
justification of earlier or more recent experimental observations for
semiclassical behavior in simple-connected space (when the particles pass
through full magnetic fields), and also new extensions to more general cases
of delocalized (spread-out) quantum states, (ii) a nontrivial correction to
misleading or even incorrect results that appear often in the literature (both
for static and time-dependent cases), and (iii) a natural remedy for Causality
\textquotedblleft paradoxes\textquotedblright\ in time-dependent Aharonov-Bohm
configurations. These nonlocal contributions seem to have escaped from
state-of-the-art path-integral approaches. An extension of the method applied
to the fields (rather than the \textquotedblleft gauge
function\textquotedblright\ $\Lambda$) indicates that these nonlocalities
demonstrate a causal propagation of phases of quantum mechanical wavefunctions
(and these can possibly address causal issues in time-dependent single- vs
double-slit experiments, an area that seems to have recently attracted
considerable interest\cite{Tollaksen},\cite{He}).

\section{Example of Generalized Solutions in Static Cases}

\bigskip By way of an example we immediately provide a simple result that will
be found later (in Section IX) for a static $(x,y)$-case (and for
simple-connected space) that generalizes the standard Dirac phase
(\ref{usualABLambda}), namely
\begin{equation}
\Lambda(x,y)=\Lambda(x_{0},y_{0})+\int_{x_{0}}^{x}A_{x}(x^{\prime
},y)dx^{\prime}+\int_{y_{0}}^{y}A_{y}(\mathbf{x}_{\mathbf{0}},y^{\prime
})dy^{\prime}+\left\{
%TCIMACRO{\dint \limits_{y_{0}}^{y}}%
%BeginExpansion
{\displaystyle\int\limits_{y_{0}}^{y}}
%EndExpansion
dy^{\prime}%
%TCIMACRO{\dint \limits_{x_{0}}^{x}}%
%BeginExpansion
{\displaystyle\int\limits_{x_{0}}^{x}}
%EndExpansion
dx^{\prime}B_{z}(x^{\prime},y^{\prime})+g(x)\right\}  \label{static1}%
\end{equation}

\[
with\text{ \ }g(x)\text{ \ }chosen\text{ \ }so\text{ \ }that\ \ \ \left\{
%TCIMACRO{\dint \limits_{y_{0}}^{y}}%
%BeginExpansion
{\displaystyle\int\limits_{y_{0}}^{y}}
%EndExpansion
dy^{\prime}%
%TCIMACRO{\dint \limits_{x_{0}}^{x}}%
%BeginExpansion
{\displaystyle\int\limits_{x_{0}}^{x}}
%EndExpansion
dx^{\prime}B_{z}(x^{\prime},y^{\prime})+g(x)\right\}  \boldsymbol{:}\text{ is
}\mathsf{independent\ of\ }\ x.
\]

In the above $B_{z}={\huge (}\boldsymbol{B}_{2}-\boldsymbol{B}_{1}%
{\huge )}_{z}$ is the difference of perpendicular magnetic fields in the two
systems, which can be nonvanishing at remote regions (see below)). The reader
should note that the first 3 terms of (\ref{static1}) are the Dirac phase
(\ref{usualABLambda}) along two perpendicular segments that connect the
initial point $(x_{0},y_{0})$ to the point of observation $(x,y)$, \textit{in
a clockwise sense }(see for example the red-arrow paths in Fig.1(b)). But
apart from this Dirac phase, we also have nonlocal contributions from $B_{z}$
and its flux within the \textquotedblleft observation
rectangle\textquotedblright\ (see i.e. the rectangle being formed by the red-
and green-arrow paths in Fig.1(b)). Below we will directly verify that
(\ref{static1}) is indeed a solution of (\ref{usualgrad}) (even for
$B_{z}(x^{\prime},y^{\prime})\neq0$ for $(x^{\prime},y^{\prime})\neq(x,y)$),
i.e. of the system \bigskip of Partial Differential Equations (PDEs)%

\begin{equation}
\frac{\partial\Lambda(x,y)}{\partial x}=A_{x}(x,y)\qquad and\qquad
\frac{\partial\Lambda(x,y)}{\partial y}=A_{y}(x,y). \label{usualgradcomps}%
\end{equation}
(Although the former is trivially satisfied (at least for cases where
interchanges of integrals with derivatives are legitimate), for the latter to
be verified one needs to simply substitute $\frac{\partial A_{x}(x^{\prime
},y)}{\partial y}$ with \ $\frac{\partial A_{y}(x^{\prime},y)}{\partial
x^{\prime}}-B_{z}(x^{\prime},y)$ \ and then carry out the integration with
respect to $x^{\prime}$ -- the reader should note the crucial appearance (and
proper placement) of $\mathbf{x}_{\mathbf{0}}\boldsymbol{\ }$ in
(\ref{static1}) for the verification of both (\ref{usualgradcomps})). It
should be noted again that (\ref{static1}) satisfies (\ref{usualgradcomps})
even for nonzero $B_{z}$ (i.e. when the particle passes through nonzero
magnetic fields in remote regions), in contradistinction to the standard
result (\ref{usualABLambda}). (For the benefit of the reader we clearly
provide in the next Section all the steps for the direct verification of
(\ref{static1})).

\bigskip

Equivalently, we will later obtain the result%

\begin{equation}
\Lambda(x,y)=\Lambda(x_{0},y_{0})+\int_{x_{0}}^{x}A_{x}(x^{\prime}%
,\mathbf{y}_{\mathbf{0}})dx^{\prime}+\int_{y_{0}}^{y}A_{y}(x,y^{\prime
})dy^{\prime}+\left\{  {\Huge -}%
%TCIMACRO{\dint \limits_{x_{0}}^{x}}%
%BeginExpansion
{\displaystyle\int\limits_{x_{0}}^{x}}
%EndExpansion
dx^{\prime}%
%TCIMACRO{\dint \limits_{y_{0}}^{y}}%
%BeginExpansion
{\displaystyle\int\limits_{y_{0}}^{y}}
%EndExpansion
dy^{\prime}B_{z}(x^{\prime},y^{\prime})+h(y)\right\}  \label{static2}%
\end{equation}%
\[
with\text{ \ }h(y)\text{ \ }chosen\text{ \ }so\text{ \ }that\text{
\ \ }\left\{  {\Huge -}%
%TCIMACRO{\dint \limits_{x_{0}}^{x}}%
%BeginExpansion
{\displaystyle\int\limits_{x_{0}}^{x}}
%EndExpansion
dx^{\prime}%
%TCIMACRO{\dint \limits_{y_{0}}^{y}}%
%BeginExpansion
{\displaystyle\int\limits_{y_{0}}^{y}}
%EndExpansion
dy^{\prime}B_{z}(x^{\prime},y^{\prime})+h(y)\right\}  \boldsymbol{:}\text{ is
}\mathsf{independent\ of\ }\ y,
\]
and again the reader should note that, apart from the first 3 terms (the Dirac
phase (\ref{usualABLambda}) along the two other (alternative) perpendicular
segments (connecting $(x_{0},y_{0})$ to $(x,y)$), now \textit{in a
counterclockwise sense }(the green-arrow paths in Fig.1(b))), we also have
nonlocal contributions from the flux of $B_{z}$ that is enclosed within the
same \textquotedblleft observation rectangle\textquotedblright\ (that is
naturally defined by the four segments of the two solutions (Fig.1(b))). It
can also be easily verified that (\ref{static2}) also satisfies the system
(\ref{usualgradcomps}) (for this $\frac{\partial A_{y}(x,y^{\prime})}{\partial
x}$ \ needs to be substituted with $\frac{\partial A_{x}(x,y^{\prime}%
)}{\partial y^{\prime}}+B_{z}(x,y^{\prime})$ \ and then integration with
respect to $y^{\prime}$ needs to be carried out, with the proper appearance
(and placement)\ of \ $\mathbf{y}_{\mathbf{0}}$ in (\ref{static2}) now being
the crucial element $-$ see direct verification in the next Section).

In all the above, $A_{x}$ and $A_{y}$ are the Cartesian components of
$\ \boldsymbol{A}\mathbf{(r)}=\boldsymbol{A}(x,y)=\boldsymbol{A}%
_{\boldsymbol{2}}(\mathbf{r})-\boldsymbol{A}_{\boldsymbol{1}}(\mathbf{r})$,
and, as already mentioned, \ $B_{z}$ \ is the difference between
(perpendicular) magnetic fields that the two systems may experience in regions
that \textit{do not contain} the observation point $(x,y)$ (i.e.
$B_{z}(x^{\prime},y^{\prime})={\huge (}\boldsymbol{B}_{\boldsymbol{2}%
}(x^{\prime},y^{\prime})-\boldsymbol{B}_{\boldsymbol{1}}(x^{\prime},y^{\prime
}){\huge )}_{z}=\frac{\partial A_{y}(x^{\prime},y^{\prime})}{\partial
x^{\prime}}-$ $\frac{\partial A_{x}(x^{\prime},y^{\prime})}{\partial
y^{\prime}}$, \ which can be nonzero for \ $(x^{\prime},y^{\prime})\neq(x,y)$).

In the present and following Section we place the emphasis in pointing out
(and proving) the new solutions (that apparently have been overlooked in the
literature). In later Sections, we will see that these results actually
demonstrate that the passage of particles through magnetic fields has the
effect of cancelling Aharonov-Bohm types of phases. And in the special case of
semiclassical motion we will suggest an understanding of this cancellation in
terms of the experimentally observed compatibility (or consistency) of
Aharonov-Bohm fringe-displacement and trajectory-deflection due to the Lorentz
force. (The corresponding \textquotedblleft electric analog\textquotedblright%
\ of this consistency of trajectory-behavior will also be pointed out).
However, the above cancellations are true even for completely delocalized
states (and the deeper reason for this will be obvious from the derivation of
the above two solutions $-$ the origin of the cancellations being essentially
the single-valuedness of phases for simple-connected space). Therefore,
generalized results such as the above go beyond the usual Aharonov-Bohm
behaviors reviewed in the Introductory Sections, and give an extended
description of physical systems in more complex physical arrangements. [It is
simply mentioned here that cancellations of the above type will be extended
and generalized further to cases that also involve the time variable $t$;
these will be presented in later Sections, with a detailed description of how
they are derived. Interpreted in a different way, such cancellations $-$
through the new nonlocal terms $-$ will take away the \textquotedblleft
mystery\textquotedblright\ of why certain classical arguments (based on past
history and the Faraday's law of Induction) seem to \textquotedblleft
work\textquotedblright\ (give the correct Aharonov-Bohm phases in static
arrangements, by invoking the \textit{history} of how the experimental set up
was built at earlier times). Although more generally useful, some simple first
applications of these results will also be given that provide a natural remedy
for well-known \textquotedblleft paradoxes\textquotedblright\ in
time-dependent Aharonov-Bohm configurations, and are indicative of an even
more general causal propagation of wavefunction phases in the Schr\"{o}dinger picture].

\section{\textbf{Elementary\ Verification\ Of\ Above Solutions (even for cases
with }$\boldsymbol{B}_{z}\boldsymbol{\neq0}$\textbf{)}}

In static cases, and simple-connected space, let us call our solution
(\ref{static1}) $\Lambda_{1}$, namely%

\[
\Lambda_{1}(x,y)=\Lambda_{1}(x_{0},y_{0})+%
%TCIMACRO{\dint \limits_{x_{0}}^{x}}%
%BeginExpansion
{\displaystyle\int\limits_{x_{0}}^{x}}
%EndExpansion
A_{x}(x^{\prime},y)dx^{\prime}+%
%TCIMACRO{\dint \limits_{y_{0}}^{y}}%
%BeginExpansion
{\displaystyle\int\limits_{y_{0}}^{y}}
%EndExpansion
A_{y}(x_{0},y^{\prime})dy^{\prime}+\left\{
%TCIMACRO{\dint \limits_{y_{0}}^{y}}%
%BeginExpansion
{\displaystyle\int\limits_{y_{0}}^{y}}
%EndExpansion
dy^{\prime}%
%TCIMACRO{\dint \limits_{x_{0}}^{x}}%
%BeginExpansion
{\displaystyle\int\limits_{x_{0}}^{x}}
%EndExpansion
dx^{\prime}B_{z}(x^{\prime},y^{\prime})+g(x)\right\}
\]
with\ $g(x)$ chosen so that\ $\ \left\{
%TCIMACRO{\dint \limits^{y}}%
%BeginExpansion
{\displaystyle\int\limits^{y}}
%EndExpansion%
%TCIMACRO{\dint \limits^{x}}%
%BeginExpansion
{\displaystyle\int\limits^{x}}
%EndExpansion
B_{z}+g(x)\right\}  \boldsymbol{\ }$is independent of $x.$

Verification that it solves the system of PDEs (\ref{usualgradcomps}) (even
for $B_{z}(x^{\prime},y^{\prime})\neq0$)$\boldsymbol{:}$

\textbf{A) \ }$\frac{\partial\Lambda_{1}(x,y)}{\partial x}=A_{x}(x,y)\qquad
$satisfied trivially\qquad$\checkmark$

(because $\left\{  ...\right\}  $ is independent of $x$).

\textbf{B) \ }$\frac{\partial\Lambda_{1}(x,y)}{\partial y}=%
%TCIMACRO{\dint \limits_{x_{0}}^{x}}%
%BeginExpansion
{\displaystyle\int\limits_{x_{0}}^{x}}
%EndExpansion
\frac{\partial A_{x}(x^{\prime},y)}{\partial y}dx^{\prime}+A_{y}(x_{0},y)+%
%TCIMACRO{\dint \limits_{x_{0}}^{x}}%
%BeginExpansion
{\displaystyle\int\limits_{x_{0}}^{x}}
%EndExpansion
B_{z}(x^{\prime},y)dx^{\prime}+\frac{\partial g(x)}{\partial y},$

(the last term being trivially zero, $\frac{\partial g(x)}{\partial y}=0$),
and then with the substitution

$\frac{\partial A_{x}(x^{\prime},y)}{\partial y}=\frac{\partial A_{y}%
(x^{\prime},y)}{\partial x^{\prime}}-B_{z}(x^{\prime},y)$

we obtain

$\frac{\partial\Lambda_{1}(x,y)}{\partial y}=%
%TCIMACRO{\dint \limits_{x_{0}}^{x}}%
%BeginExpansion
{\displaystyle\int\limits_{x_{0}}^{x}}
%EndExpansion
\frac{\partial A_{y}(x^{\prime},y)}{\partial x^{\prime}}dx^{\prime}-%
%TCIMACRO{\dint \limits_{x_{0}}^{x}}%
%BeginExpansion
{\displaystyle\int\limits_{x_{0}}^{x}}
%EndExpansion
B_{z}(x^{\prime},y)dx^{\prime}+A_{y}(x_{0},y)+%
%TCIMACRO{\dint \limits_{x_{0}}^{x}}%
%BeginExpansion
{\displaystyle\int\limits_{x_{0}}^{x}}
%EndExpansion
B_{z}(x^{\prime},y)dx^{\prime}.$

(i) We see that the 2nd and 4th terms of the right-hand-side (rhs)
\textit{cancel each other}, and

(ii) the 1st term of the rhs is $%
%TCIMACRO{\dint \limits_{x_{0}}^{x}}%
%BeginExpansion
{\displaystyle\int\limits_{x_{0}}^{x}}
%EndExpansion
\frac{\partial A_{y}(x^{\prime},y)}{\partial x^{\prime}}dx^{\prime}%
=A_{y}(x,y)-A_{y}(x_{0},y).$

Hence finally

$\frac{\partial\Lambda_{1}(x,y)}{\partial y}=A_{y}(x,y).\qquad\checkmark$

\bigskip

We have directly shown therefore that the basic system of PDEs
(\ref{usualgradcomps}) is indeed satisfied by our \textbf{generalized}
solution $\Lambda_{1}(x,y),$ \textbf{even for any nonzero} $B_{z}(x^{\prime
},y^{\prime})$\ (in regions $(x^{\prime},y^{\prime})\neq(x,y)$).

In a completely analogous way, one can easily see that our alternative
solution (eqn.(\ref{static2})) also satisfies the basic system of PDEs above.
Below we give the direct proof$\boldsymbol{:}$

Let us call our second static solution (eqn.(\ref{static2})) $\Lambda_{2}$, namely%

\[
\Lambda_{2}(x,y)=\Lambda_{2}(x_{0},y_{0})+\int_{x_{0}}^{x}A_{x}(x^{\prime
},y_{0})dx^{\prime}+\int_{y_{0}}^{y}A_{y}(x,y^{\prime})dy^{\prime}+\left\{
{\Huge -}%
%TCIMACRO{\dint \limits_{x_{0}}^{x}}%
%BeginExpansion
{\displaystyle\int\limits_{x_{0}}^{x}}
%EndExpansion
dx^{\prime}%
%TCIMACRO{\dint \limits_{y_{0}}^{y}}%
%BeginExpansion
{\displaystyle\int\limits_{y_{0}}^{y}}
%EndExpansion
dy^{\prime}B_{z}(x^{\prime},y^{\prime})+h(y)\right\}
\]
with $h(y)$ chosen so that\ $\ \left\{  {\Huge -}%
%TCIMACRO{\dint \limits^{x}}%
%BeginExpansion
{\displaystyle\int\limits^{x}}
%EndExpansion%
%TCIMACRO{\dint \limits^{y}}%
%BeginExpansion
{\displaystyle\int\limits^{y}}
%EndExpansion
B_{z}+h(y)\right\}  \boldsymbol{\ }$is independent of\textsf{ }$y$.

\bigskip

Verification that it solves the system of PDEs (\ref{usualgradcomps}) (even
for $B_{z}(x^{\prime},y^{\prime})\neq0$)$\boldsymbol{:}$

\bigskip

\textbf{A) \ }$\frac{\partial\Lambda_{2}(x,y)}{\partial y}=A_{y}(x,y)\qquad
$satisfied trivially\qquad$\checkmark$

(because $\left\{  ...\right\}  $ is independent of $y$).

\bigskip

\textbf{B) \ }$\frac{\partial\Lambda_{2}(x,y)}{\partial x}=A_{x}(x,y_{0})+%
%TCIMACRO{\dint \limits_{y_{0}}^{y}}%
%BeginExpansion
{\displaystyle\int\limits_{y_{0}}^{y}}
%EndExpansion
\frac{\partial A_{y}(x,y^{\prime})}{\partial x}dy^{\prime}-%
%TCIMACRO{\dint \limits_{y_{0}}^{y}}%
%BeginExpansion
{\displaystyle\int\limits_{y_{0}}^{y}}
%EndExpansion
B_{z}(x,y^{\prime})dy^{\prime}+\frac{\partial h(y)}{\partial x},$

(the last term being trivially zero, $\frac{\partial h(y)}{\partial x}=0$),
and then with the substitution

$\frac{\partial A_{y}(x,y^{\prime})}{\partial x}=\frac{\partial A_{x}%
(x,y^{\prime})}{\partial y^{\prime}}+B_{z}(x,y^{\prime})$

we obtain

$\frac{\partial\Lambda_{2}(x,y)}{\partial x}=A_{x}(x,y_{0})+%
%TCIMACRO{\dint \limits_{y_{0}}^{y}}%
%BeginExpansion
{\displaystyle\int\limits_{y_{0}}^{y}}
%EndExpansion
\frac{\partial A_{x}(x,y^{\prime})}{\partial y^{\prime}}dy^{\prime}+%
%TCIMACRO{\dint \limits_{y_{0}}^{y}}%
%BeginExpansion
{\displaystyle\int\limits_{y_{0}}^{y}}
%EndExpansion
B_{z}(x,y^{\prime})dy^{\prime}-%
%TCIMACRO{\dint \limits_{y_{0}}^{y}}%
%BeginExpansion
{\displaystyle\int\limits_{y_{0}}^{y}}
%EndExpansion
B_{z}(x,y^{\prime})dy^{\prime}.$

(i) We see that the last two terms of the rhs \textit{cancel each other}, and

(ii) the 2nd term of the rhs is $%
%TCIMACRO{\dint \limits_{y_{0}}^{y}}%
%BeginExpansion
{\displaystyle\int\limits_{y_{0}}^{y}}
%EndExpansion
\frac{\partial A_{x}(x,y^{\prime})}{\partial y^{\prime}}dy^{\prime}%
=A_{x}(x,y)-A_{x}(x,y_{0}).$

Hence finally

$\frac{\partial\Lambda_{2}(x,y)}{\partial x}=A_{x}(x,y).\qquad\checkmark$

\bigskip Once again, all the above are true for any nonzero $B_{z}(x^{\prime
},y^{\prime})$ (in regions $(x^{\prime},y^{\prime})\neq(x,y)$).

\section{Simple Examples$\boldsymbol{:}$ New results shown in explicit form}

To see how the above solutions appear in nontrivial cases (and how they give
completely new results, i.e. \textbf{not differing from the usual ones (i.e.
from the Dirac phase) by a mere constant}) let us first take examples of
striped $B_{z}$-distributions in spacetime$\boldsymbol{:}$

\textbf{(a)} For the case of an extended \textit{vertical} strip - parallel to
the $y$-axis, such as in Fig.1(a) (with $t$ replaced by $y$) (i.e. the
particle has actually passed through nonzero $B_{z}$, hence through
\textit{different} magnetic fields in the two (mapped) systems), then, for $x$
located outside (and on the right of) the strip, the quantity$\
%TCIMACRO{\dint \limits_{y_{0}}^{y}}%
%BeginExpansion
{\displaystyle\int\limits_{y_{0}}^{y}}
%EndExpansion
dy^{\prime}%
%TCIMACRO{\dint \limits_{x_{0}}^{x}}%
%BeginExpansion
{\displaystyle\int\limits_{x_{0}}^{x}}
%EndExpansion
dx^{\prime}B_{z}(x^{\prime},y^{\prime})$ in\ $\Lambda_{1}$ is already
independent of$\ x$\ (since a displacement of the $(x,y)$-corner of the
rectangle to the right, along the $x$-direction, does not change the enclosed
magnetic flux $-$ see Fig. 1(a) for the analogous $(x,t)$-case that will be
discussed in following Sections)$\boldsymbol{;}$ hence in this case the
function $g(x)$ can be taken as $g(x)=0$ (up to a constant $C$) and the
condition for $g(x)$ stated in the solution eqn.(\ref{static1}) (i.e. that the
quantity in brackets must be independent of $x$) is indeed satisfied.

So for this setup, the nonlocal term in the solution \textbf{survives} (the
quantity in brackets is nonvanishing), but \textbf{it is not constant}%
$\boldsymbol{:}$ this enclosed flux depends on $y$ (since the enclosed flux
\textbf{does change} with a displacement of the $(x,y)$-corner of the
rectangle upwards, along the $y$-direction). Hence, by looking at the
alternative solution $\Lambda_{2}(x,y),$ the quantity$\
%TCIMACRO{\dint \limits_{x_{0}}^{x}}%
%BeginExpansion
{\displaystyle\int\limits_{x_{0}}^{x}}
%EndExpansion
dx^{\prime}%
%TCIMACRO{\dint \limits_{y_{0}}^{y}}%
%BeginExpansion
{\displaystyle\int\limits_{y_{0}}^{y}}
%EndExpansion
dy^{\prime}B_{z}(x^{\prime},y^{\prime})$\ is dependent on$\ y$, so that $h(y)$
must be chosen as $\ h(y)=+%
%TCIMACRO{\dint \limits_{x_{0}}^{x}}%
%BeginExpansion
{\displaystyle\int\limits_{x_{0}}^{x}}
%EndExpansion
dx^{\prime}%
%TCIMACRO{\dint \limits_{y_{0}}^{y}}%
%BeginExpansion
{\displaystyle\int\limits_{y_{0}}^{y}}
%EndExpansion
dy^{\prime}B_{z}(x^{\prime},y^{\prime})$ (up to the same constant $C$)\ in
order to \textit{cancel} the $y$-dependence, so that its own condition stated
in the solution eqn.(\ref{static2}) (i.e. that the quantity in brackets must
be independent of $y$) is satisfied$\boldsymbol{;}$ as a result, the quantity
in brackets in solution $\Lambda_{2}$ disappears and there is no nonlocal
contribution in $\Lambda_{2}$ (for $C=0$). (Of course, if we had used a
$C\neq0$, the nonlocal contributions would be distributed between the two
solutions in a different manner, but without changing the Physics when we take
the \textit{difference} of the two solutions (see below)).

With these choices of $h(y)$ and $g(x)$, we already have new results (compared
to the standard ones of the integrals of potentials). I.e. one of the two
solutions, namely $\Lambda_{1}$ \textbf{is} affected nonlocally by the
enclosed flux (and this flux is \textit{not} constant). Spelled out clearly,
the two results are:%

\[
\Lambda_{1}(x,y)=\Lambda_{1}(x_{0},y_{0})+%
%TCIMACRO{\dint \limits_{x_{0}}^{x}}%
%BeginExpansion
{\displaystyle\int\limits_{x_{0}}^{x}}
%EndExpansion
A_{x}(x^{\prime},y)dx^{\prime}+%
%TCIMACRO{\dint \limits_{y_{0}}^{y}}%
%BeginExpansion
{\displaystyle\int\limits_{y_{0}}^{y}}
%EndExpansion
A_{y}(x_{0},y^{\prime})dy^{\prime}+%
%TCIMACRO{\dint \limits_{y_{0}}^{y}}%
%BeginExpansion
{\displaystyle\int\limits_{y_{0}}^{y}}
%EndExpansion
dy^{\prime}%
%TCIMACRO{\dint \limits_{x_{0}}^{x}}%
%BeginExpansion
{\displaystyle\int\limits_{x_{0}}^{x}}
%EndExpansion
dx^{\prime}B_{z}(x^{\prime},y^{\prime})+C
\]

\[
\Lambda_{2}(x,y)=\Lambda_{2}(x_{0},y_{0})+\int_{x_{0}}^{x}A_{x}(x^{\prime
},y_{0})dx^{\prime}+\int_{y_{0}}^{y}A_{y}(x,y^{\prime})dy^{\prime}+C.
\]
And it is easy to see that, if we subtract the two solutions $\Lambda_{1}$ and
$\Lambda_{2}$, the result is \textit{zero} (because the line integrals of the
vector potential $\boldsymbol{A}$ in the two solutions are in opposite senses
in the $(x,y)$ plane, hence their difference leads to a \textit{closed} line
integral of $\boldsymbol{A}$ which is in turn equal to the enclosed magnetic
flux, and this flux always happens to be of opposite sign from that of the
enclosed flux that explicitly appears through the nonlocal\ term of the
$B_{z}$-field that survives in $\Lambda_{1}$)$\boldsymbol{.}$ (In the above we
of course assumed single-valuedness of $\Lambda$ at the initial point
$(x_{0},y_{0})$, i.e. $\Lambda_{1}(x_{0},y_{0})=\Lambda_{2}(x_{0},y_{0});$
matters of multivaluedness of $\Lambda$ at the observation point $(x,y)$ will
be addressed later, in Section IX).

The reader should probably note that the above equality of the two solutions
is due to the fact that the $x$-independent quantity in brackets of the 1st
solution (\ref{static1}) is equal to the function $h(y)$ of the 2nd solution
(\ref{static2}), and the $y$-independent quantity in brackets of the 2nd
solution (\ref{static2}) is equal to the function $g(x)$ of the first solution
(\ref{static1}). This will turn out to be a general behavioral pattern of the
two solutions in simple-connected space, that will be valid for any shape of
$B_{z}$-distribution, as will be shown later.

This vanishing of $\Lambda_{1}(x,y)-\Lambda_{2}(x,y)$ is a cancellation effect
that is emphasized further (and generally proved) later below (and can be
viewed as a generalization of the Werner \& Brill experimental
observations\cite{WernerBrill} to even completely delocalized states, as will
be fully discussed in physical terms in Section IX). It basically originates
from the single-valuedness of $\Lambda$ at $(x,y)$ for simple-connected space.
This effect is generalized even further in later Sections (i.e. also to cases
of combined 3 variables $x,y,t$) for the van Kampen
thought-experiment\cite{vanKampen} (where we will have a combination of
spatial multiple-connectivity at an initial instant $t_{0}$, and
simple-connectivity in $(x,t)$ and $(y,t)$ planes).

\textbf{(b) }In the \textquotedblleft dual case\textquotedblright\ of an
extended \textit{horizontal} strip - parallel to the $x$-axis, the proper
choices (for $y$ above the strip) are basically reverse (i.e. we can now take
$h(y)=0$ \ and $g(x)=-%
%TCIMACRO{\dint \limits_{y_{0}}^{y}}%
%BeginExpansion
{\displaystyle\int\limits_{y_{0}}^{y}}
%EndExpansion
dy^{\prime}%
%TCIMACRO{\dint \limits_{x_{0}}^{x}}%
%BeginExpansion
{\displaystyle\int\limits_{x_{0}}^{x}}
%EndExpansion
dx^{\prime}B_{z}(x^{\prime},y^{\prime})$ $\ $(since the flux enclosed in the
rectangle now depends on $x$, but not on $y$), with both choices always up to
a common constant) and once again we can easily see a similar cancellation
effect. In this case again, the results are new (a \ nonlocal term now
surviving in $\Lambda_{2}$). Again spelled out clearly, these are:%

\[
\Lambda_{1}(x,y)=\Lambda_{1}(x_{0},y_{0})+%
%TCIMACRO{\dint \limits_{x_{0}}^{x}}%
%BeginExpansion
{\displaystyle\int\limits_{x_{0}}^{x}}
%EndExpansion
A_{x}(x^{\prime},y)dx^{\prime}+%
%TCIMACRO{\dint \limits_{y_{0}}^{y}}%
%BeginExpansion
{\displaystyle\int\limits_{y_{0}}^{y}}
%EndExpansion
A_{y}(x_{0},y^{\prime})dy^{\prime}+C
\]

\[
\Lambda_{2}(x,y)=\Lambda_{2}(x_{0},y_{0})+\int_{x_{0}}^{x}A_{x}(x^{\prime
},y_{0})dx^{\prime}+\int_{y_{0}}^{y}A_{y}(x,y^{\prime})dy^{\prime}{\Huge -}%
%TCIMACRO{\dint \limits_{x_{0}}^{x}}%
%BeginExpansion
{\displaystyle\int\limits_{x_{0}}^{x}}
%EndExpansion
dx^{\prime}%
%TCIMACRO{\dint \limits_{y_{0}}^{y}}%
%BeginExpansion
{\displaystyle\int\limits_{y_{0}}^{y}}
%EndExpansion
dy^{\prime}B_{z}(x^{\prime},y^{\prime})+C
\]
(their difference also being zero -- a generalized Werner \& Brill
cancellation (see Section IX for further discussion)).

\bigskip

\textbf{(c) }If we want cases that are more involved (i.e. with the nonlocal
contributions appearing nontrivially in \textbf{both} solutions $\Lambda_{1}$
and $\Lambda_{2}$ and with $g(x)$ and $h(y)$ not being \textquotedblleft
immediately visible\textquotedblright), we must consider different shapes of
$B_{z}$-distributions. One such case is a triangular one that is shown in
Fig.1(b) (for simplicity an equilateral triangle, with the initial point
$(x_{0},y_{0})=(0,0)$) and with the point of observation $(x,y)$ being fairly
close to the triangle's right side as in the Figure. Note that for such a
configuration, the part of the magnetic flux that is inside the
\textquotedblleft observation rectangle\textquotedblright\ (defined by the
right upper corner $(x,y)$) depends on \textbf{both} $x$ \textbf{and} $y$. It
turns out, however, that this $(x$ and $y)-$dependent enclosed flux can be
written as a sum of separate $x$- and $y$-contributions, so that appropriate
$g(x)$ and $h(y)$ can be found (each one of them must be chosen so that it
only cancels the corresponding variable's dependence of the enclosed flux).
For a homogeneous $B_{z}$ it is a rather straightforward exercise to determine
this enclosed part, i.e. the common area between the observation rectangle and
the equilateral triangle, and from this we can find the appropriate $g(x)$
that will cancel the $x$-dependence, and the appropriate $h(y)$ that will
cancel the $y$-dependence. These appropriate choices turn out to be

\begin{equation}
g(x)=B_{z}\left[  \mathbf{-(}\sqrt{3}ax-\frac{\sqrt{3}}{2}x^{2})+\frac
{\sqrt{3}}{4}a^{2}\right]  +C \label{triangular1}%
\end{equation}
and%

\begin{equation}
h(y)=B_{z}\left[  \mathbf{(}ay-\frac{y^{2}}{\sqrt{3}})-\frac{\sqrt{3}}{4}%
a^{2}\right]  +C \label{triangular2}%
\end{equation}
with $a$ being the side of the equilateral triangle. (We again note that a
physical arbitrariness described by the common constant $C$, does not play any
role when we take the difference of the two solutions (\ref{static1}) and
(\ref{static2})). We should emphasize that the above results, if combined with
(\ref{static1}) or (\ref{static2}), give the nontrivial nonlocal contributions
of the difference $B_{z}$ of the remote magnetic fields on $\Lambda$ of each
solution (hence on the phase of the wavefunction of each wavepacket travelling
along each path) at the observation point $(x,y)$. (We mention again that in
the case of completely spread-out states, the equality of the two solutions at
the observation point essentially demonstrates the uniqueness
(single-valuedness) of the phase in simple-connected space). Further physical
discussion, and a semiclassical interpretation is given later, in Section IX
and in the Final Sections of the paper.

In more \textquotedblleft difficult\textquotedblright\ geometries, i.e. when
the shape of the $B_{z}$-distribution is such that the enclosed flux does
\textbf{not }decouple in a sum of separate $x$- and $y$-contributions, \ such
as cases of circularly shaped distributions, it is advantageous to solve the
system (\ref{usualgrad}) directly in non-Cartesian (i.e. polar) coordinates.
This is done further below in Section IX.

Finally, the reader may wonder how the usual magnetic Aharonov-Bohm effect
arises in the above formulation, and here is probably the best place to
provide an explanation (although we will need for this to invoke the most
general results -- for multiple-connected space -- that will be derived
later). For the Aharonov-Bohm setting we will have to deal with
multiple-connected space and with a (static) magnetic flux $\Phi$ being
contained only in the physically inaccessible region.\ In such a case we know
that the $\Lambda(\mathbf{r})$ that solves (\ref{usualgrad}) is not
single-valued. How is this fact (and the standard result (\ref{usualABLambda}%
)) compatible with the new formulation? To answer this in full generality we
will consider two separate cases that arise naturally (pertaining to the issue
of what the dummy variables $(x^{\prime},y^{\prime})$ inside the $B_{z}$-terms
of our results (i.e. of (\ref{static1}) and (\ref{static2})) actually
represent). First, if the variables $x$ and $y$ everywhere above always denote
only coordinates of the region that is physically accessible to the particle,
then $B_{z}$ above is everywhere vanishing, this effectively reducing
(\ref{static1}) and (\ref{static2}) to%

\[
\Lambda(x,y)=\Lambda(x_{0},y_{0})+\int_{x_{0}}^{x}A_{x}(x^{\prime
},y)dx^{\prime}+\int_{y_{0}}^{y}A_{y}(\mathbf{x}_{\mathbf{0}},y^{\prime
})dy^{\prime}+C
\]

\[
\Lambda(x,y)=\Lambda(x_{0},y_{0})+\int_{x_{0}}^{x}A_{x}(x^{\prime}%
,\mathbf{y}_{\mathbf{0}})dx^{\prime}+\int_{y_{0}}^{y}A_{y}(x,y^{\prime
})dy^{\prime}+C
\]
with $C$ a common constant; these are the standard results (the Dirac phases)
along the two alternative paths discussed above (the red and green paths of
Fig.1) that (through their difference) lead to the magnetic Aharonov-Bohm
effect ($\Lambda$ being no longer single-valued and the difference of the two
solutions giving the enclosed (and physically inaccessible) $\Phi).$ Let us
however be even more general and let us decide to use the variables $x$ and
$y$ to \textit{also} denote coordinates of the physically inaccessible
region$\boldsymbol{;}$ this would be the case, if, for example, we had
previously started with that region being accessible (i.e. through a
penetrable scalar potential) and at the end we followed a limiting procedure
(i.e. of this scalar potential going to infinity) so that this region would
become in the limit impenetrable and therefore inaccessible. In such a case
the variables $x$ and $y$ would now contain \textit{remnants} of the
previously allowed values (but currently not allowed for the description of
particle coordinates) such as the values of the dummy variables $x^{\prime}$
and $y^{\prime}$ in the $B_{z}$-terms of (\ref{static1}) and (\ref{static2}%
)$\boldsymbol{;}$ such values would therefore still be present in the
expressions giving $\Lambda$ (even though these dummy variables $x^{\prime}$
and $y^{\prime}$ would now describe an inaccessible region). In other words,
the inaccessible $B_{z}$ is still formally present in the problem and it shows
up explicitly in the generalized gauge functions of the new formulation. How
does this formulation then lead to the standard Aharonov-Bohm result in such a
limiting case (essentially a case of smoothly-induced spatial multiple-connectivity)?

Before we answer this, the reader should probably be reminded that our
formulation only deals with wavefunction-phases$\boldsymbol{;}$ questions
therefore of rigid (vanishing) boundary conditions (on the boundary of the
inaccessible region) that apply to (and must be imposed on) the entire
wavefunction, and mostly on its modulus, can only be addressed indirectly (and
as we will see, through a \textquotedblleft memory\textquotedblright\ that the
phases have of their multivaluedness, whenever the space is
multiple-connected). To see this, we need two slightly generalized results
that will be rigorously derived later (eqns (\ref{Lambda(x,y)1}) and
(\ref{Lambda(x,y)4})) that add certain constants (what we will later call
\textquotedblleft multiplicities\textquotedblright) to the above
\textquotedblleft simple-connected\textquotedblright\ forms (\ref{static1})
and (\ref{static2}). These most general results (for multiple-connected space)
will be derived in Section IX and will turn out to be%

\[
\Lambda(x,y)=\Lambda(x_{0},y_{0})+%
%TCIMACRO{\dint \limits_{x_{0}}^{x}}%
%BeginExpansion
{\displaystyle\int\limits_{x_{0}}^{x}}
%EndExpansion
A_{x}(x^{\prime},y)dx^{\prime}+%
%TCIMACRO{\dint \limits_{y_{0}}^{y}}%
%BeginExpansion
{\displaystyle\int\limits_{y_{0}}^{y}}
%EndExpansion
A_{y}(x_{0},y^{\prime})dy^{\prime}+\left\{
%TCIMACRO{\dint \limits_{y_{0}}^{y}}%
%BeginExpansion
{\displaystyle\int\limits_{y_{0}}^{y}}
%EndExpansion
dy^{\prime}%
%TCIMACRO{\dint \limits_{x_{0}}^{x}}%
%BeginExpansion
{\displaystyle\int\limits_{x_{0}}^{x}}
%EndExpansion
dx^{\prime}B_{z}(x^{\prime},y^{\prime})+g(x)\right\}  +f(y_{0})
\]
and%
\[
\Lambda(x,y)=\Lambda(x_{0},y_{0})+\int_{x_{0}}^{x}A_{x}(x^{\prime}%
,y_{0})dx^{\prime}+\int_{y_{0}}^{y}A_{y}(x,y^{\prime})dy^{\prime}+\left\{
{\Huge -}%
%TCIMACRO{\dint \limits_{x_{0}}^{x}}%
%BeginExpansion
{\displaystyle\int\limits_{x_{0}}^{x}}
%EndExpansion
dx^{\prime}%
%TCIMACRO{\dint \limits_{y_{0}}^{y}}%
%BeginExpansion
{\displaystyle\int\limits_{y_{0}}^{y}}
%EndExpansion
dy^{\prime}B_{z}(x^{\prime},y^{\prime})+h(y)\right\}  +\hat{h}(x_{0})
\]
with the functions $g(x)$ and $h(y)$ satisfying the same conditions as in
(\ref{static1}) and (\ref{static2}). We note the extra appearance of the new
constant terms $f(y_{0})$ and $\hat{h}(x_{0})$ (the \textquotedblleft
multiplicities\textquotedblright) and these are \textquotedblleft
defined\textquotedblright\ (see (\ref{1stintegration(x,y)}) and
(\ref{2ndintegration(x,y)}) where the functions $f$ and $\hat{h}$ will be
first introduced) by%
\[
f(y_{0})=\Lambda(x,y_{0})-\Lambda(x_{0},y_{0})-\int_{x_{0}}^{x}A_{x}%
(x^{\prime},y_{0})dx^{\prime}%
\]
and%
\[
\hat{h}(x_{0})=\Lambda(x_{0},y)-\Lambda(x_{0},y_{0})-\int_{y_{0}}^{y}%
A_{y}(x_{0},y^{\prime})dy^{\prime}.
\]
Let us then identify proper choices for the functions $g(x)$ and $h(y)$ and
for the constants $f(y_{0})$ and $\hat{h}(x_{0})$ in the above case of spatial
multiple-connectivity (such as the standard magnetic Aharonov-Bohm case, with
a non-extended (and static) magnetic flux in the forbidden
region)$\boldsymbol{:}$ First, we can always take $g(x)=0$ and\ $h(y)=0$
(always up to a common constant as discussed earlier), since the enclosed
magnetic flux is (in this Aharonov-Bohm case) independent of both $x$ and $y$
-- the conditions of $g(x)$ and $h(y)$ being then automatically satisfied.
Second, let us look more closely at the above \textquotedblleft
definitions\textquotedblright\ of $f(y_{0})$ and $\hat{h}(x_{0})\boldsymbol{:}%
$ we first note that $f(y_{0})$ must be independent of $x$, and this is indeed
true as is apparent by formally taking the derivative of the above definition
of $f(y_{0})$ with respect to $x\boldsymbol{;}$ we then have $\frac{\partial
f(y_{0})}{\partial x}=$ $\frac{\partial\Lambda(x,y_{0})}{\partial x}%
-A_{x}(x,y_{0})$ which is indeed zero (as $\Lambda(x,y)$ satisfies by
assumption the first equation of the system (\ref{usualgradcomps}) of PDEs
(evaluated at $y=y_{0}$)), showing that $\frac{\partial f(y_{0})}{\partial
x}=0$ and that $f(y_{0})$ does not really depend on the variable $x$ that
appears in its definition. We can therefore determine its value by taking the
limit $x\rightarrow x_{0}$ (for fixed $y_{0}$)$\boldsymbol{:}$ we see from the
above that this limit is simply equal to $\lim_{x\rightarrow x_{0}}%
\Lambda(x,y_{0})-\Lambda(x_{0},y_{0})$ [we leave out cases where $A_{x}$ has a
$\delta$-function form, as will be discussed later in the careful derivations
of all our results where interchanges of integrals must be allowed], and this
difference is nonzero only when there is a multivaluedness of $\Lambda$ at the
point $(x_{0},y_{0})$, as \textit{is }actually our case. The limit
$x\rightarrow x_{0}$ (for fixed $y_{0}$) in the path-sense of solution
(\ref{static1}) (or of (\ref{Lambda(x,y)1})) that is then needed here in order
to determine $f(y_{0})$, is equivalent to making an entire closed trip around
the observation rectangle in the \textit{negative} sense, landing on the
initial point $(x_{0},y_{0}),$ this therefore giving the value $f(y_{0})=$
minus enclosed magnetic flux $=-\Phi$ (which is indeed a constant independent
of $x$ and $y$, as it should be).\ By following a completely symmetric
argument for the above definition of $\hat{h}(x_{0})$ (and by now taking the
limit $y\rightarrow y_{0}$ (for fixed $x_{0}$), that is now equivalent to
going around the loop in the \textit{positive} sense, landing again on the
initial point $(x_{0},y_{0})$) we obtain\ $\hat{h}(x_{0})=+\Phi$. If these
values of $f(y_{0})$ and $\hat{h}(x_{0})$ are finally substituted in the above
most general solutions (eqns (\ref{Lambda(x,y)1}) and (\ref{Lambda(x,y)4}))
together with $g(x)=h(y)=0$, then we note that $f(y_{0})$ cancels out the
{\Huge \ }$%
%TCIMACRO{\dint \limits_{y_{0}}^{y}}%
%BeginExpansion
{\displaystyle\int\limits_{y_{0}}^{y}}
%EndExpansion
dy^{\prime}%
%TCIMACRO{\dint \limits_{x_{0}}^{x}}%
%BeginExpansion
{\displaystyle\int\limits_{x_{0}}^{x}}
%EndExpansion
dx^{\prime}B_{z}(x^{\prime},y^{\prime})$ term (which is here just equal to the
inaccessible flux $\Phi$), and $\hat{h}(x_{0})$ cancels out the \ {\Huge -}$%
%TCIMACRO{\dint \limits_{x_{0}}^{x}}%
%BeginExpansion
{\displaystyle\int\limits_{x_{0}}^{x}}
%EndExpansion
dx^{\prime}%
%TCIMACRO{\dint \limits_{y_{0}}^{y}}%
%BeginExpansion
{\displaystyle\int\limits_{y_{0}}^{y}}
%EndExpansion
dy^{\prime}B_{z}(x^{\prime},y^{\prime})$ term, and the two solutions are then
once again reduced to the usual solutions of mere $A$-integrals along the two
paths (i.e. the standard Dirac phase, with no nonlocal contributions) -- their
difference giving once again the closed loop integral of $\boldsymbol{A},$
hence the inaccessible flux and, finally, the well-known magnetic
Aharonov-Bohm result. One should note again the expected, namely that the
standard result in the new formulation requires some effort and it is only
derived indirectly (due to the fact that we only deal with phases and not the
moduli of wavefunctions, on which boundary conditions are normally imposed),
and it basically comes from the \textquotedblleft memory\textquotedblright\ of
the multivaluedness that the \textquotedblleft gauge
function\textquotedblright\ $\Lambda$ carries (due to the
multiple-connectivity of space).

\section{Example of Generalized Solutions in Dynamic Cases, with full
derivation}

Let us now look at a case with full time-dependence. Although it is now
probably easy for the reader to guess the corresponding generalized results,
i.e. for a spatially-one-dimensional $(x,t)-$problem (i.e. by Euclidean
rotation (in 4-D spacetime) from the above solutions), we nevertheless start
from the beginning and give a full physical discussion $-$ as this \textit{is}
the case that actually led us to the above generalized solutions, and a case
associated with a number of misleading arguments (and often incorrect results)
propagating in the literature.

Let us then first focus on the simplest case of one-dimensional quantum
systems, i.e. a single quantum particle of charge $\ q$, but in the presence
of the most general (spatially nonuniform and time-dependent) vector and
scalar potentials, and ask the following question$\boldsymbol{:}$ what is the
gauge function $\Lambda(x,t)$ that takes us from (maps) a system with
potentials $A_{1}(x,t)$ and $\phi_{1}(x,t)$ to a system with potentials
$A_{2}(x,t)$ and $\phi_{2}(x,t)$? (meaning the usual mapping (\ref{UsualPsi})
between the wavefunctions of the two systems through the phase factor
$\frac{q}{\hbar c}\Lambda(x,t)$). [Of course for this mapping to be possible
we assume that at the point $(x,t)$ of observation (or \textquotedblleft
measurement\textquotedblright\ of $\Lambda$ or the wavefunction $\Psi$) \ we
have equal electric fields ($E_{i}=-\nabla\phi_{i}-\frac{1}{c}\frac{\partial
A_{i}}{\partial t}$), namely

\begin{equation}
-\frac{\partial\phi_{2}(x,t)}{\partial x}-\frac{1}{c}\frac{\partial
A_{2}(x,t)}{\partial t}=-\frac{\partial\phi_{1}(x,t)}{\partial x}-\frac{1}%
{c}\frac{\partial A_{1}(x,t)}{\partial t} \label{Efield}%
\end{equation}
(so that the $A$'s and $\phi$'s in (\ref{Efield}) can satisfy the basic system
of equations (\ref{gaugetransf}), or equivalently, of the system of equations
(\ref{xt-BasicSystem}) below), but we will \textit{not} exclude the
possibility of the two systems passing through \textit{different }electric
fields in different regions of spacetime, i.e. for $(x^{\prime},t^{\prime
})\neq(x,t)$. In fact, this possibility \textbf{will come out naturally} from
a careful solution of the basic system (\ref{xt-BasicSystem})$\boldsymbol{;}$
it is for example straightforward for the reader to immediately verify that
the results (\ref{LambdaStatic1}) or (\ref{LambdaStatic2}) that will be
derived below (and will contain contributions of electric fields from remote
regions of spacetime) indeed satisfy the basic input system of equations
(\ref{xt-BasicSystem}), something that will be explicitly verified in the next Section].

Returning to the question on the appropriate $\Lambda$ that takes us from the
set $(A_{1},\phi_{1})$ to the set $(A_{2},\phi_{2})$, we note that, in cases
of static vector potentials ($A(x)$'s) \textit{and} spatially uniform scalar
potentials ($\phi(t)$'s) the answer usually given is the well-known%

\begin{equation}
\Lambda(x,t)=\Lambda(x_{0},t_{0})+\int_{x_{0}}^{x}A(x^{\prime})dx^{\prime
}-c\int_{t_{0}}^{t}\phi(t^{\prime})dt^{\prime} \label{LambdaUsual}%
\end{equation}
with $\ A(x)=A_{2}(x)-A_{1}(x)$ \ and $\ \phi(t)=\phi_{2}(t)-\phi_{1}(t)$ (and
it can be viewed as a combination of (\ref{usualABLambda}) and
(\ref{usualelectricABLambda}), being immediately applicable to the description
of cases of \textit{combined} magnetic and electric Aharonov-Bohm effects
reviewed in the Introductory Sections).

\bigskip

In the most general case (and with the variables$\ x$\ and$\ t$ being
\textbf{completely uncorrelated}), it is often stated in the literature [e.g.
in Brown \& Holland\cite{BrownHolland}, see i.e. their eqn. (57) for vanishing
boost velocity $\mathbf{v}=0$] that the appropriate $\Lambda$ has a form that
is a plausible extention of (\ref{LambdaUsual}), namely%

\begin{equation}
\Lambda(x,t)=\Lambda(x_{0},t_{0})+%
%TCIMACRO{\dint \limits_{x_{0}}^{x}}%
%BeginExpansion
{\displaystyle\int\limits_{x_{0}}^{x}}
%EndExpansion
\left[  A_{2}(x^{\prime},t)-A_{1}(x^{\prime},t)\right]  dx^{\prime}-c%
%TCIMACRO{\dint \limits_{t_{0}}^{t}}%
%BeginExpansion
{\displaystyle\int\limits_{t_{0}}^{t}}
%EndExpansion
\left[  \phi_{2}(x,t^{\prime})-\phi_{1}(x,t^{\prime})\right]  dt^{\prime}.
\label{BrownHolland}%
\end{equation}

This form is certainly \ \textit{incorrect} \ for uncorrelated variables $x$
and $t$ \ (the reader can easily verify that the system of equations
(\ref{xt-BasicSystem}) below is ${\large not}$ satisfied by
(\ref{BrownHolland})$\boldsymbol{;}$ indeed$\boldsymbol{:}$ (i) When the
$\frac{\partial}{\partial x}$ operator acts on eq.(\ref{BrownHolland}), it
gives the correct $A(x,t)$ from the 1st term, but it also gives some annoying
additional nonzero quantity from the 2nd term (that survives because of the
$x$-dependence of $\phi$); hence it invalidates the first of the basic system
of PDEs. (ii) Similarly, when the $\mathbf{-}\frac{1}{c}\frac{\partial
}{\partial t}$ operator acts on eq.(\ref{BrownHolland}), it gives the correct
$\phi(x,t)$ from the 2nd term, but it also gives some annoying additional
nonzero quantity from the 1st term (that survives because of the
$t$-dependence of $A$); hence it invalidates the second of the basic system of
PDEs. It is only when $A$ is $t$-independent, and $\phi$ is
spatially-independent, that eq.(\ref{BrownHolland}) is correct. It is also
interesting to note that the line integrals appearing in (\ref{BrownHolland})
do not form a path (in spacetime) that contains the initial to the final point
(see below)). [An alternative form that is also given in the literature is
again eq.(\ref{BrownHolland}), but with the variables that are not integrated
over implicitly assumed to belong to the initial point (hence a $t_{0}$
replaces $t$ in $A$, and an $x_{0}$ replaces $x$ in $\phi$). However, one can
see again that the basic system of PDEs is not satisfied (the above
differential operators, when acted on $\Lambda$, give $A(x,t_{0})$ and
$\phi(x_{0},t)$, hence not the values of the potentials at the point of
observation $(x,t)$ as they should), this not being an acceptable solution
either. And in this case also there is no spacetime-path connecting the
initial $(x_{0},t_{0})$ to the final point $(x,t)$ either]. In the present
work we will find that the correct form consists of two terms$\boldsymbol{:}$
one is rather trivial (and leads to the natural appearance of a \textit{path}
that connects initial and final points in spacetime, a property that
(\ref{BrownHolland}) does \textit{not} have (see eqns.(\ref{LambdaStatic1})
and (\ref{LambdaStatic2}) below for the corrected \textquotedblleft
path-forms\textquotedblright\ in the line integrals of potentials)), but the
second term is nontrivial$\boldsymbol{:}$ it consists of nonlocal
contributions of classical electric fields from remote regions of space-time.
We will discuss below the consequences of these terms and we will later show
that such nonlocal contributions also appear (in an extended form) in more
general situations, i.e. they are also present in higher spatial
dimensionality (and they then also involve remote magnetic fields in
combination with the electric ones)$\boldsymbol{;}$ these lead to
modifications of ordinary Aharonov-Bohm behaviors or have other consequences,
one of them being a natural remedy of Causality \textquotedblleft
paradoxes\textquotedblright\ in time-dependent Aharonov-Bohm experiments. (An
application of the method to the integral forms of Maxwell's equations will
also be briefly mentioned, which, although not the main focus of this paper,
gives an important causal interpretation of these temporal nonlocalities of
wavefunction phases in the general case).

\bigskip

The form (\ref{BrownHolland}) commonly used is of course motivated by the
well-known Wu \& Yang\cite{WuYang} nonintegrable phase factor, that has a
phase equal to \ $\int A_{\mu}dx^{\mu}=\int Adx-c\int\phi dt$, \ a form that
appears naturally within\ the framework of path-integral treatments, or
generally in physical situations where narrow wavepackets are implicitly
assumed for the quantum particle\textbf{:} the integrals appearing in
(\ref{BrownHolland}) are then taken along particle trajectories (hence spatial
and temporal variables \textit{not} being uncorrelated, but being connected in
a particular manner to produce the path$\boldsymbol{;}$ all integrals are
therefore basically only time-integrals). But even then, eqn.
(\ref{BrownHolland}) is valid only when these trajectories are always (in time
and in space) inside identical classical fields for the two (mapped) systems.
Here, however, we will be focusing on what a canonical (and not a
path-integral or other semiclassical) treatment gives us$\boldsymbol{;}$ this
will cover the general case of arbitrary wavefunctions that can even be
completely delocalized, and will also allow the particle to travel through
different classical fields for the two systems in remote spacetime regions
(i.e. \ $E_{2}\left(  x,t^{\prime}\right)  \neq E_{1}\left(  x,t\right)  $
\ if $\ \ t^{\prime}<t$ \ etc.).

\bigskip

It is therefore clear that in order to find the appropriate $\ \Lambda(x,t)$
\ that answers the above question in full generality will require a careful
solution of the system of PDEs (\ref{gaugetransf}), applied to only one
spatial variable, namely%

\begin{equation}
\frac{\partial\Lambda(x,t)}{\partial x}=A(x,t)\qquad and\qquad-\frac{1}%
{c}\frac{\partial\Lambda(x,t)}{\partial t}=\phi\left(  x,t\right)
\label{xt-BasicSystem}%
\end{equation}
(with $\ A(x,t)=A_{2}(x,t)-A_{1}(x,t)$ \ and $\ \phi\left(  x,t\right)
=\phi_{2}\left(  x,t\right)  -\phi_{1}\left(  x,t\right)  $), the system being
underdetermined in the sense that we only have knowledge of $\Lambda$ at an
initial point $(x_{0},t_{0})$ and with no further boundary conditions (hence
multiplicities of solutions being generally expected, and these are discussed
separately below). Let us first look for unique (single-valued) solutions
(i.e. with $\Lambda$ being a \textit{function }on the $(x,t)$-plane\textit{,}
in the sense of elementary analysis) and let us integrate the\textit{\ first}
of (\ref{xt-BasicSystem}) -- without dropping terms that may at first sight
appear redundant -- to obtain%

\begin{equation}
\Lambda(x,t)-\Lambda(x_{0},t)=\int_{x_{0}}^{x}A(x^{\prime},t)dx^{\prime}%
+\tau(t). \label{1stintegration}%
\end{equation}

By then substituting this to the second of (\ref{xt-BasicSystem}) (and
assuming that interchanges of derivatives and integrals are allowed, i.e.
covering cases of potentials with discontinuous first derivatives, something
that corresponds to the physical case of discontinuous magnetic fields - a
case very often discussed in the literature), we obtain%

\begin{equation}
\phi\left(  x,t\right)  =-\frac{1}{c}%
%TCIMACRO{\dint \limits_{x_{0}}^{x}}%
%BeginExpansion
{\displaystyle\int\limits_{x_{0}}^{x}}
%EndExpansion
\frac{\partial A(x^{\prime},t)}{\partial t}dx^{\prime}-\frac{1}{c}%
\frac{\partial\tau(t)}{\partial t}-\frac{1}{c}\frac{\partial\Lambda(x_{0}%
,t)}{\partial t}, \label{Phi}%
\end{equation}
which if integrated gives%

\begin{equation}
\tau(t)=\tau(t_{0})+\Lambda(x_{0},t_{0})-\Lambda(x_{0},t)-%
%TCIMACRO{\dint \limits_{t_{0}}^{t}}%
%BeginExpansion
{\displaystyle\int\limits_{t_{0}}^{t}}
%EndExpansion
dt^{\prime}%
%TCIMACRO{\dint \limits_{x_{0}}^{x}}%
%BeginExpansion
{\displaystyle\int\limits_{x_{0}}^{x}}
%EndExpansion
dx^{\prime}\frac{\partial A(x^{\prime},t^{\prime})}{\partial t^{\prime}}%
-c\int_{t_{0}}^{t}\phi\left(  x,t^{\prime}\right)  dt^{\prime}+g(x)
\label{Tau}%
\end{equation}
with $g(x)$ to be chosen in such a way that the entire right-hand-side of
(\ref{Tau}) is only a function of$\ t$ (hence independent of$\ x$). Finally,
by substituting \ \ $\frac{\partial A(x^{\prime},t^{\prime})}{\partial
t^{\prime}}$ \ with \ \ $-c\left(  E(x^{\prime},t^{\prime})+\frac{\partial
\phi(x^{\prime},t^{\prime})}{\partial x^{\prime}}\right)  $, \ (where
$E(x^{\prime},t^{\prime})=E_{2}(x^{\prime},t^{\prime})-E_{1}(x^{\prime
},t^{\prime})$), carrying out the integration with respect to $x^{\prime}$,
and by demanding that $\tau(t)$ \ be independent of $\ x$, we finally obtain
the following general solution%

\begin{equation}
\Lambda(x,t)=\Lambda(x_{0},t_{0})+%
%TCIMACRO{\dint \limits_{x_{0}}^{x}}%
%BeginExpansion
{\displaystyle\int\limits_{x_{0}}^{x}}
%EndExpansion
A(x^{\prime},t)dx^{\prime}-c%
%TCIMACRO{\dint \limits_{t_{0}}^{t}}%
%BeginExpansion
{\displaystyle\int\limits_{t_{0}}^{t}}
%EndExpansion
\phi(x_{0},t^{\prime})dt^{\prime}+\left\{  c%
%TCIMACRO{\dint \limits_{t_{0}}^{t}}%
%BeginExpansion
{\displaystyle\int\limits_{t_{0}}^{t}}
%EndExpansion
dt^{\prime}%
%TCIMACRO{\dint \limits_{x_{0}}^{x}}%
%BeginExpansion
{\displaystyle\int\limits_{x_{0}}^{x}}
%EndExpansion
dx^{\prime}E(x^{\prime},t^{\prime})+g(x)\right\}  +\tau(t_{0})
\label{LambdaStatic1}%
\end{equation}
with $\ g(x)$ \ chosen in such a way that the quantity $\ \ \left\{  c%
%TCIMACRO{\dint \limits_{t_{0}}^{t}}%
%BeginExpansion
{\displaystyle\int\limits_{t_{0}}^{t}}
%EndExpansion
dt^{\prime}%
%TCIMACRO{\dint \limits_{x_{0}}^{x}}%
%BeginExpansion
{\displaystyle\int\limits_{x_{0}}^{x}}
%EndExpansion
dx^{\prime}E(x^{\prime},t^{\prime})+g(x)\right\}  $ \ \ is independent of $x$.

\bigskip

Here it should be noted that, if we had first integrated the \textit{second}
of (\ref{xt-BasicSystem}) we would have%

\begin{equation}
\Lambda(x,t)-\Lambda(x,t_{0})=-c\int_{t_{0}}^{t}\phi(x,t^{\prime})dt^{\prime
}+\chi(x) \label{2ndintegration}%
\end{equation}

and then from the first of (\ref{xt-BasicSystem}) we would get%

\begin{equation}
A\left(  x,t\right)  =-c%
%TCIMACRO{\dint \limits_{t_{0}}^{t}}%
%BeginExpansion
{\displaystyle\int\limits_{t_{0}}^{t}}
%EndExpansion
\frac{\partial\phi(x,t^{\prime})}{\partial x}dt^{\prime}+\frac{\partial
\chi(x)}{\partial x}+\frac{\partial\Lambda(x,t_{0})}{\partial x},
\label{Aintermediate}%
\end{equation}

which after integration would give%

\begin{equation}
\chi(x)=\chi(x_{0})+\Lambda(x_{0},t_{0})-\Lambda(x,t_{0})+c%
%TCIMACRO{\dint \limits_{x_{0}}^{x}}%
%BeginExpansion
{\displaystyle\int\limits_{x_{0}}^{x}}
%EndExpansion
dx^{\prime}%
%TCIMACRO{\dint \limits_{t_{0}}^{t}}%
%BeginExpansion
{\displaystyle\int\limits_{t_{0}}^{t}}
%EndExpansion
dt^{\prime}\frac{\partial\phi(x^{\prime},t^{\prime})}{\partial x^{\prime}%
}+\int_{x_{0}}^{x}A\left(  x^{\prime},t\right)  dx^{\prime}+\hat{g}(t)
\label{chi}%
\end{equation}
with $\hat{g}(t)$ to be chosen in such a way that the entire right-hand-side
of (\ref{chi}) is only a function of$\ x$ (hence independent of$\ t$).
Finally, by substituting \ \ $\frac{\partial\phi(x^{\prime},t^{\prime}%
)}{\partial x^{\prime}}$ \ with \ \ $-\left(  E(x^{\prime},t^{\prime}%
)+\frac{1}{c}\frac{\partial A(x^{\prime},t^{\prime})}{\partial t^{\prime}%
}\right)  $, carrying out the integration with respect to $t^{\prime}$, and by
demanding that $\chi(x)$ \ be independent of $\ t$, we would finally obtain
the following general solution%

\begin{equation}
\Lambda(x,t)=\Lambda(x_{0},t_{0})+%
%TCIMACRO{\dint \limits_{x_{0}}^{x}}%
%BeginExpansion
{\displaystyle\int\limits_{x_{0}}^{x}}
%EndExpansion
A(x^{\prime},t_{0})dx^{\prime}-c\int_{t_{0}}^{t}\phi\left(  x,t^{\prime
}\right)  dt^{\prime}+\left\{  -c%
%TCIMACRO{\dint \limits_{x_{0}}^{x}}%
%BeginExpansion
{\displaystyle\int\limits_{x_{0}}^{x}}
%EndExpansion
dx^{\prime}%
%TCIMACRO{\dint \limits_{t_{0}}^{t}}%
%BeginExpansion
{\displaystyle\int\limits_{t_{0}}^{t}}
%EndExpansion
dt^{\prime}E(x^{\prime},t^{\prime})+\hat{g}(t)\right\}  +\chi(x_{0})
\label{LambdaStatic2}%
\end{equation}
with $\ \hat{g}(t)$ \ chosen in such a way that the quantity \ $\ \left\{  -c%
%TCIMACRO{\dint \limits_{x_{0}}^{x}}%
%BeginExpansion
{\displaystyle\int\limits_{x_{0}}^{x}}
%EndExpansion
dx^{\prime}%
%TCIMACRO{\dint \limits_{t_{0}}^{t}}%
%BeginExpansion
{\displaystyle\int\limits_{t_{0}}^{t}}
%EndExpansion
dt^{\prime}E(x^{\prime},t^{\prime})+\hat{g}(t)\right\}  \ \ $\ \ is
independent of$\ t$.

\bigskip Solutions (\ref{LambdaStatic1}) and (\ref{LambdaStatic2}) can be
viewed as the (formal) analogs of (\ref{static1}) and (\ref{static2})
correspondingly, although they hide in them much richer Physics because of
their dynamic character (see Section VIII). (The additional constant last
terms will be shown in Section VIII to be related to possible multiplicities
of $\Lambda$, and they are zero in simple-connected spacetimes).

\bigskip The reader is once again provided with the direct verification that
(\ref{LambdaStatic1}) or (\ref{LambdaStatic2}) are indeed solutions of the
basic system of PDEs (\ref{xt-BasicSystem}) in the Section that follows.

\section{Verification of solutions and simple dynamical examples}

Let us call our first solution \ (eqn.(\ref{LambdaStatic1})) for
simple-connected spacetime $\Lambda_{3}$, namely%

\[
\Lambda_{3}(x,t)=\Lambda_{3}(x_{0},t_{0})+%
%TCIMACRO{\dint \limits_{x_{0}}^{x}}%
%BeginExpansion
{\displaystyle\int\limits_{x_{0}}^{x}}
%EndExpansion
A(x^{\prime},t)dx^{\prime}-c%
%TCIMACRO{\dint \limits_{t_{0}}^{t}}%
%BeginExpansion
{\displaystyle\int\limits_{t_{0}}^{t}}
%EndExpansion
\phi(x_{0},t^{\prime})dt^{\prime}+\left\{  c%
%TCIMACRO{\dint \limits_{t_{0}}^{t}}%
%BeginExpansion
{\displaystyle\int\limits_{t_{0}}^{t}}
%EndExpansion
dt^{\prime}%
%TCIMACRO{\dint \limits_{x_{0}}^{x}}%
%BeginExpansion
{\displaystyle\int\limits_{x_{0}}^{x}}
%EndExpansion
dx^{\prime}E(x^{\prime},t^{\prime})+g(x)\right\}
\]

with $g(x)$ chosen so that $\left\{  c%
%TCIMACRO{\dint \limits_{t_{0}}^{t}}%
%BeginExpansion
{\displaystyle\int\limits_{t_{0}}^{t}}
%EndExpansion
dt^{\prime}%
%TCIMACRO{\dint \limits_{x_{0}}^{x}}%
%BeginExpansion
{\displaystyle\int\limits_{x_{0}}^{x}}
%EndExpansion
dx^{\prime}E(x^{\prime},t^{\prime})+g(x)\right\}  $ is independent of $x$.

Verification that it solves the system of PDEs (\ref{xt-BasicSystem}) (even
for $E(x^{\prime},t^{\prime})\neq0$)$\boldsymbol{:}$

\bigskip

\textbf{A)} $\ \frac{\partial\Lambda_{3}(x,t)}{\partial x}=A(x,t)\qquad
$satisfied trivially\qquad$\checkmark$ \ \ \ \ \ \ \ \ \ \ \ 

(because $\left\{  ....\right\}  $ is independent of $x$).

\bigskip

\bigskip\textbf{B)} \ $-\frac{1}{c}\frac{\partial\Lambda_{3}(x,t)}{\partial
t}=-\frac{1}{c}%
%TCIMACRO{\dint \limits_{x_{0}}^{x}}%
%BeginExpansion
{\displaystyle\int\limits_{x_{0}}^{x}}
%EndExpansion
\frac{\partial A(x^{\prime},t)}{\partial t}dx^{\prime}+\phi(x_{0},t)-%
%TCIMACRO{\dint \limits_{x_{0}}^{x}}%
%BeginExpansion
{\displaystyle\int\limits_{x_{0}}^{x}}
%EndExpansion
E(x^{\prime},t)dx^{\prime}-\frac{1}{c}\frac{\partial g(x)}{\partial t}$
,\ \ \ \ \qquad

(the last term being trivially zero, $\frac{\partial g(x)}{\partial t}=0$ ),
and then with the substitution

$-\frac{1}{c}\frac{\partial A(x^{\prime},t)}{\partial t}=\frac{\partial
\phi(x^{\prime},t)}{\partial x^{\prime}}+E(x^{\prime},t)$

we obtain

$-\frac{1}{c}\frac{\partial\Lambda_{3}(x,t)}{\partial t}=%
%TCIMACRO{\dint \limits_{x_{0}}^{x}}%
%BeginExpansion
{\displaystyle\int\limits_{x_{0}}^{x}}
%EndExpansion
\frac{\partial\phi(x^{\prime},t)}{\partial x^{\prime}}dx^{\prime}+%
%TCIMACRO{\dint \limits_{x_{0}}^{x}}%
%BeginExpansion
{\displaystyle\int\limits_{x_{0}}^{x}}
%EndExpansion
E(x^{\prime},t)dx^{\prime}+\phi(x_{0},t)-%
%TCIMACRO{\dint \limits_{x_{0}}^{x}}%
%BeginExpansion
{\displaystyle\int\limits_{x_{0}}^{x}}
%EndExpansion
E(x^{\prime},t)dx^{\prime}$. \ \ \ 

(i) We see that the 2nd and 4th terms of the rhs \textit{cancel each other}, and

(ii) the 1st term of the rhs is \ $%
%TCIMACRO{\dint \limits_{x_{0}}^{x}}%
%BeginExpansion
{\displaystyle\int\limits_{x_{0}}^{x}}
%EndExpansion
\frac{\partial\phi(x^{\prime},t)}{\partial x^{\prime}}dx^{\prime}%
=\phi(x,t)-\phi(x_{0},t).\qquad$\ 

Hence finally

$-\frac{1}{c}\frac{\partial\Lambda_{3}(x,t)}{\partial t}=\phi(x,t).\qquad
\checkmark$

\bigskip

We have directly shown therefore that the basic system of PDEs
(\ref{xt-BasicSystem}) is indeed satisfied by our \textbf{generalized}
solution $\Lambda_{3}(x,t),$ \textbf{even for any nonzero} $E(x^{\prime
},t^{\prime})$ \ (in regions $(x^{\prime},t^{\prime})\neq(x,t)$).

In a completely analogous way, one can easily see that our alternative
solution (eqn.(\ref{LambdaStatic2})) also satisfies the basic system of PDEs
above. In case this is still not clear, here is the proof$\boldsymbol{:}$

\bigskip Let us call our second (alternative) solution
(eqn.(\ref{LambdaStatic2})) again for simple-connected spacetime $\Lambda_{4}%
$, namely%

\[
\Lambda_{4}(x,t)=\Lambda_{4}(x_{0},t_{0})+%
%TCIMACRO{\dint \limits_{x_{0}}^{x}}%
%BeginExpansion
{\displaystyle\int\limits_{x_{0}}^{x}}
%EndExpansion
A(x^{\prime},t_{0})dx^{\prime}-c\int_{t_{0}}^{t}\phi\left(  x,t^{\prime
}\right)  dt^{\prime}+\left\{  -c%
%TCIMACRO{\dint \limits_{x_{0}}^{x}}%
%BeginExpansion
{\displaystyle\int\limits_{x_{0}}^{x}}
%EndExpansion
dx^{\prime}%
%TCIMACRO{\dint \limits_{t_{0}}^{t}}%
%BeginExpansion
{\displaystyle\int\limits_{t_{0}}^{t}}
%EndExpansion
dt^{\prime}E(x^{\prime},t^{\prime})+\hat{g}(t)\right\}
\]
with $\hat{g}(t)$ chosen so that $\left\{  -c%
%TCIMACRO{\dint \limits_{x_{0}}^{x}}%
%BeginExpansion
{\displaystyle\int\limits_{x_{0}}^{x}}
%EndExpansion
dx^{\prime}%
%TCIMACRO{\dint \limits_{t_{0}}^{t}}%
%BeginExpansion
{\displaystyle\int\limits_{t_{0}}^{t}}
%EndExpansion
dt^{\prime}E(x^{\prime},t^{\prime})+\hat{g}(t)\right\}  $ is independent of
$t$.

Verification that it solves the system of PDEs (\ref{xt-BasicSystem}) (even
for $E(x^{\prime},t^{\prime})\neq0$)$\boldsymbol{:}$

\bigskip

\textbf{A)} $\ -\frac{1}{c}\frac{\partial\Lambda_{4}(x,t)}{\partial t}%
=\phi(x,t)\qquad$satisfied trivially\qquad$\checkmark$ \ \ \ \ \ \ \ \ \ \ \ 

(because $\left\{  ....\right\}  $ is independent of $t$).

\bigskip

\bigskip\textbf{B)} \ $\frac{\partial\Lambda_{4}(x,t)}{\partial x}%
=A(x,t_{0})-c%
%TCIMACRO{\dint \limits_{t_{0}}^{t}}%
%BeginExpansion
{\displaystyle\int\limits_{t_{0}}^{t}}
%EndExpansion
\frac{\partial\phi(x,t^{\prime})}{\partial x}dt^{\prime}-c%
%TCIMACRO{\dint \limits_{t_{0}}^{t}}%
%BeginExpansion
{\displaystyle\int\limits_{t_{0}}^{t}}
%EndExpansion
E(x,t^{\prime})dt^{\prime}+\frac{\partial\hat{g}(t)}{\partial x}$
,\ \ \ \ \qquad

(the last term being trivially zero, $\frac{\partial\hat{g}(t)}{\partial x}=0$
), and then with the substitution

$\frac{\partial\phi(x,t^{\prime})}{\partial x^{\prime}}=-E(x,t^{\prime}%
)-\frac{1}{c}\frac{\partial A(x,t^{\prime})}{\partial t^{\prime}}$

we obtain

$\frac{\partial\Lambda_{4}(x,t)}{\partial x}=A(x,t_{0})+c%
%TCIMACRO{\dint \limits_{t_{0}}^{t}}%
%BeginExpansion
{\displaystyle\int\limits_{t_{0}}^{t}}
%EndExpansion
E(x,t^{\prime})dt^{\prime}+%
%TCIMACRO{\dint \limits_{t_{0}}^{t}}%
%BeginExpansion
{\displaystyle\int\limits_{t_{0}}^{t}}
%EndExpansion
\frac{\partial A(x,t^{\prime})}{\partial t^{\prime}}dt^{\prime}-c%
%TCIMACRO{\dint \limits_{t_{0}}^{t}}%
%BeginExpansion
{\displaystyle\int\limits_{t_{0}}^{t}}
%EndExpansion
E(x,t^{\prime})dt^{\prime}$. \ \ \ 

(i) We see that the 2nd and 4th terms of the rhs \textit{cancel each other}, and

(ii) the 3rd term of the rhs is \ $%
%TCIMACRO{\dint \limits_{t_{0}}^{t}}%
%BeginExpansion
{\displaystyle\int\limits_{t_{0}}^{t}}
%EndExpansion
\frac{\partial A(x,t^{\prime})}{\partial t^{\prime}}dt^{\prime}%
=A(x,t)-A(x,t_{0}).\qquad$\ 

Hence finally

$\frac{\partial\Lambda_{4}(x,t)}{\partial x}=A(x,t).\qquad\checkmark$

\bigskip

Once again, all the above are true for any nonzero $E(x^{\prime},t^{\prime})$
(in regions $(x^{\prime},t^{\prime})\neq(x,t)$).

\bigskip

To see again how the above solutions appear in nontrivial cases (and how they
give new results, i.e. not differing from the usual ones by a mere constant)
let us take analogous examples of strips as earlier, but now in spacetime:

\textbf{(a)} For the case of the extended \textit{vertical} strip (parallel to
the $t$-axis) of Fig.1(a) (the case of a one-dimensional capacitor that is
(arbitrarily and variably) charged for all time), then, for $x$ located
outside (and on the right of) the capacitor, the quantity$\ c%
%TCIMACRO{\dint \limits_{t_{0}}^{t}}%
%BeginExpansion
{\displaystyle\int\limits_{t_{0}}^{t}}
%EndExpansion
dt^{\prime}%
%TCIMACRO{\dint \limits_{x_{0}}^{x}}%
%BeginExpansion
{\displaystyle\int\limits_{x_{0}}^{x}}
%EndExpansion
dx^{\prime}E(x^{\prime},t^{\prime})$ in\ $\Lambda_{3}$ is already independent
of$\ x$\ (since a displacement of the $(x,t)$-corner of the rectangle to the
right, along the $x$-direction, does not change the enclosed \textquotedblleft
electric flux\textquotedblright, see Fig.1(a))$\boldsymbol{;}$ hence in this
case the function $g(x)$ can be taken as $g(x)=0$ (up to a constant $C$) and
the condition for $g(x)$ stated in the solution eqn.(\ref{LambdaStatic1})
(i.e. that the quantity in brackets must be independent of $x$) is indeed satisfied.

So for this setup, the nonlocal term in the solution \textbf{survives} (the
quantity in brackets is nonvanishing), but \textbf{it is not constant}%
$\boldsymbol{:}$ this enclosed flux depends on $t$ (since the enclosed flux
\textbf{does change} with a displacement of the $(x,t)$-corner of the
rectangle upwards, along the $t$-direction). Hence, by looking at the
alternative solution $\Lambda_{4}(x,t),$ the quantity$\ c%
%TCIMACRO{\dint \limits_{x_{0}}^{x}}%
%BeginExpansion
{\displaystyle\int\limits_{x_{0}}^{x}}
%EndExpansion
dx^{\prime}%
%TCIMACRO{\dint \limits_{t_{0}}^{t}}%
%BeginExpansion
{\displaystyle\int\limits_{t_{0}}^{t}}
%EndExpansion
dt^{\prime}E(x^{\prime},t^{\prime})$\ is dependent on$\ t$, so that $\hat
{g}(t)$ must be chosen as $\ \hat{g}(t)=+c%
%TCIMACRO{\dint \limits_{x_{0}}^{x}}%
%BeginExpansion
{\displaystyle\int\limits_{x_{0}}^{x}}
%EndExpansion
dx^{\prime}%
%TCIMACRO{\dint \limits_{t_{0}}^{t}}%
%BeginExpansion
{\displaystyle\int\limits_{t_{0}}^{t}}
%EndExpansion
dt^{\prime}E(x^{\prime},t^{\prime})$ (up to the same constant $C$)\ in order
to \textit{cancel} the $t$-dependence, so that its own condition stated in the
solution eqn.(\ref{LambdaStatic2}) (i.e. that the quantity in brackets must be
independent of $t$) is satisfied$\boldsymbol{;}$ as a result, the quantity in
brackets in solution $\Lambda_{4}$ disappears and there is no nonlocal
contribution in $\Lambda_{4}$ (for $C=0$). (Once again, if we had used a
$C\neq0$, the nonlocal contributions would be distributed differently between
the two solutions, but again without changing the Physics when we take the
\textit{difference} of the two solutions).

With these choices of $\hat{g}(t)$ and $g(x)$, we already have new results
(compared to the standard ones of the integrals of potentials). I.e. one of
the two solutions, namely $\Lambda_{3}$ \textbf{is} affected nonlocally by the
enclosed flux (and this flux is \textit{not} constant). Spelled out clearly,
the two results are:%

\[
\Lambda_{3}(x,t)=\Lambda_{3}(x_{0},t_{0})+%
%TCIMACRO{\dint \limits_{x_{0}}^{x}}%
%BeginExpansion
{\displaystyle\int\limits_{x_{0}}^{x}}
%EndExpansion
A(x^{\prime},t)dx^{\prime}-c%
%TCIMACRO{\dint \limits_{t_{0}}^{t}}%
%BeginExpansion
{\displaystyle\int\limits_{t_{0}}^{t}}
%EndExpansion
\phi(x_{0},t^{\prime})dt^{\prime}+c%
%TCIMACRO{\dint \limits_{t_{0}}^{t}}%
%BeginExpansion
{\displaystyle\int\limits_{t_{0}}^{t}}
%EndExpansion
dt^{\prime}%
%TCIMACRO{\dint \limits_{x_{0}}^{x}}%
%BeginExpansion
{\displaystyle\int\limits_{x_{0}}^{x}}
%EndExpansion
dx^{\prime}E(x^{\prime},t^{\prime})+C
\]

\[
\Lambda_{4}(x,t)=\Lambda_{4}(x_{0},t_{0})+%
%TCIMACRO{\dint \limits_{x_{0}}^{x}}%
%BeginExpansion
{\displaystyle\int\limits_{x_{0}}^{x}}
%EndExpansion
A(x^{\prime},t_{0})dx^{\prime}-c\int_{t_{0}}^{t}\phi\left(  x,t^{\prime
}\right)  dt^{\prime}+C
\]
(and their difference, as mentioned above, is zero - denoting what might be
called a generalized Werner \& Brill cancellation in spacetime).

\bigskip

\textbf{(b) }In the \textquotedblleft dual case\textquotedblright\ of an
extended \textit{horizontal} strip - parallel to the $x$-axis (that
corresponds to a nonzero electric field in all space that has however a finite
duration $T)$, the proper choices (for observation time instant $t>T$) are
basically reverse (i.e. we can now take $\hat{g}(t)=0$ \ and $g(x)=-c%
%TCIMACRO{\dint \limits_{t_{0}}^{t}}%
%BeginExpansion
{\displaystyle\int\limits_{t_{0}}^{t}}
%EndExpansion
dt^{\prime}%
%TCIMACRO{\dint \limits_{x_{0}}^{x}}%
%BeginExpansion
{\displaystyle\int\limits_{x_{0}}^{x}}
%EndExpansion
dx^{\prime}E(x^{\prime},t^{\prime})$ (since the \textquotedblleft electric
flux\textquotedblright\ enclosed in the \textquotedblleft observation
rectangle\textquotedblright\ now depends on $x$, but not on $t$), with both
choices always up to a common constant) and once again we can easily see a
similar cancellation effect. In this case again, the results are new (a
\ nonlocal term now surviving in $\Lambda_{4}$). Again spelled out clearly,
these are:%

\[
\Lambda_{3}(x,t)=\Lambda_{3}(x_{0},t_{0})+%
%TCIMACRO{\dint \limits_{x_{0}}^{x}}%
%BeginExpansion
{\displaystyle\int\limits_{x_{0}}^{x}}
%EndExpansion
A(x^{\prime},t)dx^{\prime}-c%
%TCIMACRO{\dint \limits_{t_{0}}^{t}}%
%BeginExpansion
{\displaystyle\int\limits_{t_{0}}^{t}}
%EndExpansion
\phi(x_{0},t^{\prime})dt^{\prime}+C
\]

\[
\Lambda_{4}(x,t)=\Lambda_{4}(x_{0},t_{0})+%
%TCIMACRO{\dint \limits_{x_{0}}^{x}}%
%BeginExpansion
{\displaystyle\int\limits_{x_{0}}^{x}}
%EndExpansion
A(x^{\prime},t_{0})dx^{\prime}-c\int_{t_{0}}^{t}\phi\left(  x,t^{\prime
}\right)  dt^{\prime}-c%
%TCIMACRO{\dint \limits_{x_{0}}^{x}}%
%BeginExpansion
{\displaystyle\int\limits_{x_{0}}^{x}}
%EndExpansion
dx^{\prime}%
%TCIMACRO{\dint \limits_{t_{0}}^{t}}%
%BeginExpansion
{\displaystyle\int\limits_{t_{0}}^{t}}
%EndExpansion
dt^{\prime}E(x^{\prime},t^{\prime})+C
\]
(their difference also being zero -- a generalized Werner \& Brill
cancellation in spacetime).

\bigskip

\textbf{(c) }And again, if we want cases that are more involved (with the
nonlocal contributions appearing nontrivially in \textbf{both} solutions
$\Lambda_{3}$ and $\Lambda_{4}$ and with $g(x)$ and $\hat{g}(t)$ not being
\textquotedblleft immediately visible\textquotedblright) we must again
consider different shapes of $E$-distribution. One such case (the triangular)
was already shown in Fig.1(b) (for the magnetic case, which however is
completely analogous). For such a triangular case the choices of $g(x)$ and
$\hat{g}(t)$ will be different from the above and this will result in
different roles of the nonlocal terms (and these nontrivial results, or more
accurately, their analogs for the magnetic case, were given earlier in closed
analytical form, eqns (\ref{triangular1}) and (\ref{triangular2})). [And even
cases of curved shapes can be addressed more generally (when the shape is such
that the \textquotedblleft flux\textquotedblright\ does \textbf{not }decouple
in a sum of separate spatial and temporal contributions), i.e. by solving the
basic system of PDEs directly in polar coordinates (the results being
analogous to the ones given later for the magnetic cases, see eqns
(\ref{polar1})-(\ref{polarcond2}) below)].

The reader should note again that, in all the above examples in
simple-connected spacetime, the $x$-independent quantity in brackets of the
1st solution (\ref{LambdaStatic1}) is equal to the function $\hat{g}(t)$ of
the 2nd solution (\ref{LambdaStatic2}), and the $t$-independent quantity in
brackets of the 2nd solution (\ref{LambdaStatic2}) is equal to the function
$g(x)$ of the first solution (\ref{LambdaStatic1}). This pattern is what leads
to the above mentioned cancellations, and it is generally proved (i.e. for any
$E$-distribution in the $(x,t)$-plane) in the Section that follows.

\section{Comments on the general behavior of the $\boldsymbol{(x,t)}%
$-solutions}

Let us first summarize (and prove in generality) some of the behavioral
patterns that we saw in the above examples and then continue on other
properties (i.e. an account of multiplicities of $\Lambda$ in
multiple-connected spacetimes that we left out, which are described by the
constants $\tau(t_{0})$ and $\chi(x_{0})$). First, in (\ref{LambdaStatic1}) or
(\ref{LambdaStatic2}) note the proper appearance and placement of $x_{0}$ and
$t_{0}$ that gives a \textquotedblleft path-sense\textquotedblright\ to the
line integrals of potentials in each solution (with the path consisting of two
straight and perpendicular line segments connecting the initial point$\ (x_{0}%
,t_{0})$ with the final point $(x,t)$ for each solution). And there are
naturally two possible paths of this type that connect the initial point
$(x_{0},t_{0})$ with the final point $(x,t)$ (the solution
(\ref{LambdaStatic1}) having a clockwise and the solution (\ref{LambdaStatic2}%
) having a counterclockwise sense)$\boldsymbol{;}$ in this way a natural
\textit{observation} \textit{rectangle} is again formed (see Fig. 1(a)),
within which the enclosed \textquotedblleft electric fluxes\textquotedblright%
\ (in spacetime) appear to be crucial (showing up as nonlocal terms of
contributions of the electric field difference (recall that \ $E(x^{\prime
},t^{\prime})=E_{2}(x^{\prime},t^{\prime})-E_{1}(x^{\prime},t^{\prime})$) from
regions of time and space that are remote to the observation point $(x,t)$).
The appearance of these nonlocal terms (of the electric field difference) in
$\Lambda(x,t)$ from regions of space-time $(x^{\prime},t^{\prime})$ far from
the observation point $(x,t)$ seems to have a direct effect on the
wavefunction phases at $(x,t)$ (through the phase mapping that connects the
two quantum systems). The actual manner in which this happens is of course
determined by the nature of the functions $\ g(x)$ \ or $\ \hat{g}(t)$ --
these must be chosen in such a way that they satisfy their respective
conditions, as these are stated after (\ref{LambdaStatic1}) or
(\ref{LambdaStatic2}) respectively. We saw, for example, that if we have a
distribution of $E$ in the $(x,t)$-plane in the form of an extended
\textit{strip} parallel to the $t$-axis, the function $g(x)$ can be taken as
$g(x)=0$ (up to a constant $C$), and that $\hat{g}(t)$ must be chosen as
$\ \hat{g}(t)=+c%
%TCIMACRO{\dint \limits_{x_{0}}^{x}}%
%BeginExpansion
{\displaystyle\int\limits_{x_{0}}^{x}}
%EndExpansion
dx^{\prime}%
%TCIMACRO{\dint \limits_{t_{0}}^{t}}%
%BeginExpansion
{\displaystyle\int\limits_{t_{0}}^{t}}
%EndExpansion
dt^{\prime}E(x^{\prime},t^{\prime})$ (up to the same constant $C$)\ in order
to \textit{cancel} the $t$-dependence of the enclosed \textquotedblleft
flux\textquotedblright. We reemphasize that with these choices of $\hat{g}(t)$
and $g(x)$, it is easy to see that, if we subtract the two solutions
(\ref{LambdaStatic1}) and (\ref{LambdaStatic2}), the result is \textit{zero}
(because the line integrals of potentials $A$ and $\phi$ in the two solutions
are in opposite senses in the $(x,t)$ plane, hence their difference leads to a
\textit{closed} line integral which is in turn equal to the enclosed
\textquotedblleft electric flux\textquotedblright, and this flux always
happens to be of opposite sign from that of the enclosed flux that explicitly
appears through the nonlocal\ term of the $E$-fields that survives in
(\ref{LambdaStatic1}))$\boldsymbol{.}$ Such cancellation effects in dynamical
cases are important and will be discussed (and generalized) further in Section X.

Let us however give here a general proof of the above cancellations. By
looking first at the general structure of solutions (\ref{LambdaStatic1}) and
(\ref{LambdaStatic2}), we note that in both forms, the last constant terms
($\tau(t_{0})$ and $\chi(x_{0})$) are only present in cases where $\Lambda$ is
expected to be multivalued (this comes from the definitions of $\tau(t_{0})$
and $\chi(x_{0})$, see discussion below) and therefore these constant
quantities are nonvanishing in cases of motion only in multiple-connected
spacetimes (leading to phenomena of the electric Aharonov-Bohm type (see the
analogous discussion given earlier in Section V and later recapitulated in
Section IX, on the easier-to-follow magnetic case)). In such
multiple-connected cases these last terms are simply equal (in absolute value)
to the enclosed fluxes in regions of spacetime that are physically
inaccessible to the particle (in the electric Aharonov-Bohm setup, for
example, it turns out that $\tau(t_{0})=-\chi(x_{0})=$ enclosed
\textquotedblleft electric flux\textquotedblright\ in spacetime). Although
such cases can also be covered by our method below, let us for the moment
ignore them (set them to zero) and focus again on cases of motion in
simple-connected spacetimes. Then the two solutions (\ref{LambdaStatic1}) and
(\ref{LambdaStatic2}) are actually \textit{equal} as is shown below (and in so
doing, it is also shown that the $x$-independent (hence $t$-dependent)
quantity in brackets of the 1st solution (\ref{LambdaStatic1}) is equal to the
function $\hat{g}(t)$ of the 2nd solution (\ref{LambdaStatic2}) $-$ and the
$t$-independent (hence $x$-dependent) quantity in brackets of the 2nd solution
(\ref{LambdaStatic2}) is equal to the function $g(x)$ of the first solution
(\ref{LambdaStatic1})). Here is the proof$\boldsymbol{:}$

Since $\left\{  c%
%TCIMACRO{\dint \limits_{t_{0}}^{t}}%
%BeginExpansion
{\displaystyle\int\limits_{t_{0}}^{t}}
%EndExpansion
dt^{\prime}%
%TCIMACRO{\dint \limits_{x_{0}}^{x}}%
%BeginExpansion
{\displaystyle\int\limits_{x_{0}}^{x}}
%EndExpansion
dx^{\prime}E(x^{\prime},t^{\prime})+g(x)\right\}  $ is independent of $x$, its
$x$-derivative is zero which leads to $g^{\prime}(x)=-c%
%TCIMACRO{\dint \limits_{t_{0}}^{t}}%
%BeginExpansion
{\displaystyle\int\limits_{t_{0}}^{t}}
%EndExpansion
dt^{\prime}E(x,t^{\prime})$, with a general solution $\ \ g(x)=g(x_{0})-c%
%TCIMACRO{\dint \limits_{x_{0}}^{x}}%
%BeginExpansion
{\displaystyle\int\limits_{x_{0}}^{x}}
%EndExpansion
dx^{\prime}%
%TCIMACRO{\dint \limits_{t_{0}}^{t}}%
%BeginExpansion
{\displaystyle\int\limits_{t_{0}}^{t}}
%EndExpansion
dt^{\prime}E(x^{\prime},t^{\prime})+C(t),$ \ \ and with a $C(t)$ such that the
right-hand-side is only a function of$\ x$, hence independent of
$t\boldsymbol{;}$ but this is exactly the form of (\ref{LambdaStatic2}), if we
identify $C(t)$ with $\hat{g}(t)$ (and $g(x_{0})$ with $\chi(x_{0})$). This
can be easily seen if we note that substitution of $E(x^{\prime},t^{\prime})$
with \ $\ -\frac{\partial\phi(x^{\prime},t^{\prime})}{\partial x^{\prime}%
}-\frac{1}{c}\frac{\partial A(x^{\prime},t^{\prime})}{\partial t^{\prime}}$
\ and two integrations carried out finally interchange the forms of the 1st
solution (\ref{LambdaStatic1}) from $\left(
%TCIMACRO{\dint \limits_{x_{0}}^{x}}%
%BeginExpansion
{\displaystyle\int\limits_{x_{0}}^{x}}
%EndExpansion
A(x^{\prime},t)dx^{\prime}-c%
%TCIMACRO{\dint \limits_{t_{0}}^{t}}%
%BeginExpansion
{\displaystyle\int\limits_{t_{0}}^{t}}
%EndExpansion
\phi(x_{0},t^{\prime})dt^{\prime}\right)  $\ \ to $\left(
%TCIMACRO{\dint \limits_{x_{0}}^{x}}%
%BeginExpansion
{\displaystyle\int\limits_{x_{0}}^{x}}
%EndExpansion
A(x^{\prime},t_{0})dx^{\prime}-c%
%TCIMACRO{\dint \limits_{t_{0}}^{t}}%
%BeginExpansion
{\displaystyle\int\limits_{t_{0}}^{t}}
%EndExpansion
\phi(x,t^{\prime})dt^{\prime}\right)  $ of the 2nd solution
(\ref{LambdaStatic2}).

\bigskip The above could alternatively be proven if in (\ref{Tau}), instead of
substituting \ $\frac{\partial A(x^{\prime},t^{\prime})}{\partial t^{\prime}}$
\ in terms of the electric field difference, we had merely interchanged the
ordering of integrations in the 1st integral term. This would then immediately
take us to the 2nd solution (\ref{LambdaStatic2}), with automatically
identifying the $t$-independent (hence $x$-dependent) quantity \ $\left\{  -c%
%TCIMACRO{\dint \limits_{x_{0}}^{x}}%
%BeginExpansion
{\displaystyle\int\limits_{x_{0}}^{x}}
%EndExpansion
dx^{\prime}%
%TCIMACRO{\dint \limits_{t_{0}}^{t}}%
%BeginExpansion
{\displaystyle\int\limits_{t_{0}}^{t}}
%EndExpansion
dt^{\prime}E(x^{\prime},t^{\prime})+\hat{g}(t)\right\}  $ \ of the 2nd
solution (\ref{LambdaStatic2}) with the function$\ g(x)$ of the 1st solution
(\ref{LambdaStatic1}). (In a similar way, one can prove the identification of
the $x$-independent (hence $t$-dependent) quantity $\ \left\{  c%
%TCIMACRO{\dint \limits_{t_{0}}^{t}}%
%BeginExpansion
{\displaystyle\int\limits_{t_{0}}^{t}}
%EndExpansion
dt^{\prime}%
%TCIMACRO{\dint \limits_{x_{0}}^{x}}%
%BeginExpansion
{\displaystyle\int\limits_{x_{0}}^{x}}
%EndExpansion
dx^{\prime}E(x^{\prime},t^{\prime})+g(x)\right\}  $ \ of the 1st solution
(\ref{LambdaStatic1}) with the function $\hat{g}(t)$ of the 2nd solution
(\ref{LambdaStatic2})). Because of the above, it is straightforward to see (by
subtracting the two solutions) the mathematical reason for the occurence of
the cancellations claimed earlier, for any shape of $E$-distribution.

In spite therefore of the simplicity of the above considered 1-D system, we
are already in a position to draw certain very general conclusions on the
possible consequences of the new nonlocal terms of the electric fields
appearing in the solutions (\ref{LambdaStatic1}) and (\ref{LambdaStatic2}).
One can immediately see from the above considerations that these
temporally-nonlocal contributions$\boldsymbol{\ }$have the tendency of
cancelling the contributions from the $A$- and $\phi$-integrals. This already
gives an indication of cancellations that might happen in cases of higher
spatial dimensionality (where line-integrals of $A$'s, for example, can be
related to enclosed \textit{magnetic} fluxes). This \textit{is} actually the
case in the van Kampen thought-experiment that will be discussed later in
Section X $-$ although the cancellations there will be more delicate,
involving a balance among 3 variables, and with the actual \textit{senses} of
spatial closed line-integrals in the $(x,y)$-plane being nontrivially important.

Finally, with respect to $\tau(t_{0})$\ and $\chi(x_{0})$, let us give an
example to see why ordinarily (in simple-connectivity) they are zero, or in
the most general case (of multiple-connectivity) they are related to
physically inaccessible enclosed fluxes. Starting from (\ref{1stintegration}),
where $\tau(t)$ was first introduced, we have that%

\begin{equation}
\tau(t_{0})=\Lambda(x,t_{0})-\Lambda(x_{0},t_{0})-\int_{x_{0}}^{x}A(x^{\prime
},t_{0})dx^{\prime}, \label{Taut0}%
\end{equation}
which should be independent of $\ x$ \ (and \textit{it is} as can easily be
proven, since its $x$-derivative gives $\frac{\partial\Lambda(x,t_{0}%
)}{\partial x}-A(x,t_{0})$ \ which is zero, as $\Lambda(x,t)$ satisfies by
assumption the first equation of the system (\ref{xt-BasicSystem}) of PDEs
(evaluated at $t=t_{0}$)). We can therefore determine its value by taking the
limit $x\rightarrow x_{0}$ \ in (\ref{Taut0}), which is zero, unless there is
a multivaluedness of $\Lambda$ at the point $(x_{0},t_{0})$. This happens for
example for $A$ having a $\delta$-function form (a case however which we leave
out, otherwise the assumed interchanges might not be allowed) or in cases that
there is a \textquotedblleft memory\textquotedblright\ that the system has
multiplicities in $\Lambda$, i.e. in Aharonov-Bohm configurations (with
enclosed and inaccessible fluxes in space-time), hence the value of
$\tau(t_{0})$ being expected to be equal to the enclosed \textquotedblleft
electric flux\textquotedblright$\boldsymbol{:}$ the limit $x\rightarrow x_{0}$
(for fixed $t_{0}$) in the path sense of solution (\ref{LambdaStatic1}) is as
if we made an entire trip around the rectangle in the positive sense, landing
on the same initial point $(x_{0},t_{0})$. A similar argument applied for%

\begin{equation}
\chi(x_{0})=\Lambda(x_{0},t)-\Lambda(x_{0},t_{0})+c\int_{t_{0}}^{t}\phi
(x_{0},t^{\prime})dt^{\prime} \label{Chi0}%
\end{equation}
leads to the value of $\chi(x_{0})$ being equal to \textit{minus} the enclosed
\textquotedblleft electric flux\textquotedblright\ (a corresponding limit
$t\rightarrow t_{0}$ (for fixed $x_{0}$) in the path sense of solution
(\ref{LambdaStatic2}) is as if we made an entire trip around the rectangle in
the negative sense, landing on the same initial point $(x_{0},t_{0})$). If
these values are actually substituted in (\ref{LambdaStatic1}) (with $g(x)=0$)
and in (\ref{LambdaStatic2}) (with $\hat{g}(t)=0$) they give the correct
electric Aharonov-Bohm result (where effectively there are no nonlocal
contributions, and only the line-integrals of $A$ and $\phi$ contribute to the
phase). [The above choice $\ g(x)=\hat{g}(t)=0$ \ is made because, in this
Aharonov-Bohm case, the enclosed \textquotedblleft electric
flux\textquotedblright\ is independent of both $x$ and $t]$. (We should note
that the case of the electric Aharonov-Bohm setup, with the particles
traveling inside distinct equipotential cages with scalar potentials that last
for a finite duration, is the prototype of \textit{multiple-connectivity in
space-time}, a fact first noted by Iddings and reported by
Noerdlinger\cite{Iddings}. We will see later (Section XI) that this feature is
not present in the van Kampen thought-experiment, hence an electric
Aharonov-Bohm argument should not really be invoked in that case).

Before, however, leaving this simple $(x,t)$-case, we should finally emphasize
that this (or any other) contribution of electric fields is \textit{not}
present at the level of the basic Lagrangian, and the view holds in the
literature (see e.g. the work of Brown \& Home\cite{BrownHome}) that, because
of this absence, electric fields cannot contribute \textit{directly} to the
phase of the wavefunctions. This conclusion originates from the path-integral
approach (that is almost always followed), but, nevertheless, our present work
shows that fields \textit{do} contribute nonlocally. A more general discussion
on this issue is given in the final Section, after discussion of the van
Kampen thought-experiment, and also in relation to the path-integral work of
Troudet\cite{Troudet}.%

\[
\]

\section{\bigskip Again on the $\boldsymbol{(x,y)}$-Magnetic Case}

After having discussed fully the simple $(x,t)$-case, let us for completeness
give the analogous (Euclidean-rotated) derivation for $(x,y)$-variables and
briefly discuss the properties of the simpler static solutions, but now in
full generality (also including possible multi-valuedness of $\Lambda$ in
magnetic Aharonov-Bohm cases). We will simply need to apply the same
methodology (of solution of a system of PDEs) to such static spatially
two-dimensional cases (so that now different (remote) magnetic fields for the
two systems, perpendicular to the 2-D space, will arise). For such cases we
need to solve the system of PDEs already shown in (\ref{usualgradcomps}), namely%

\[
\frac{\partial\Lambda(x,y)}{\partial x}=A_{x}(x,y)\qquad and\qquad
\frac{\partial\Lambda(x,y)}{\partial y}=A_{y}(x,y).
\]
By first integrating the 1st of this (again without dropping any terms that
may appear redundant) we obtain the analog of (\ref{1stintegration}), namely%

\begin{equation}
\Lambda(x,y)-\Lambda(x_{0},y)=\int_{x_{0}}^{x}A_{x}(x^{\prime},y)dx^{\prime
}+f(y) \label{1stintegration(x,y)}%
\end{equation}
and by then substituting the result to the 2nd we have%

\begin{equation}
A_{y}\left(  x,y\right)  =%
%TCIMACRO{\dint \limits_{x_{0}}^{x}}%
%BeginExpansion
{\displaystyle\int\limits_{x_{0}}^{x}}
%EndExpansion
\frac{\partial A_{x}(x^{\prime},y)}{\partial y}dx^{\prime}+f^{\prime}%
(y)+\frac{\partial\Lambda(x_{0},y)}{\partial y} \label{Ay}%
\end{equation}
which if integrated leads to%

\begin{equation}
f(y)=f(y_{0})-\Lambda(x_{0},y)+\Lambda(x_{0},y_{0})-%
%TCIMACRO{\dint \limits_{y_{0}}^{y}}%
%BeginExpansion
{\displaystyle\int\limits_{y_{0}}^{y}}
%EndExpansion
dy^{\prime}%
%TCIMACRO{\dint \limits_{x_{0}}^{x}}%
%BeginExpansion
{\displaystyle\int\limits_{x_{0}}^{x}}
%EndExpansion
dx^{\prime}\frac{\partial A_{x}(x^{\prime},y^{\prime})}{\partial y^{\prime}%
}+\int_{y_{0}}^{y}A_{y}\left(  x,y^{\prime}\right)  dy^{\prime}+g(x)
\label{f(y)}%
\end{equation}
with $g(x)$ to be chosen in such a way that the entire right-hand-side of
(\ref{f(y)}) is only a function of $\ y$ (hence independent of $\ x$).
Finally, by substituting \ \ $\frac{\partial A_{x}(x^{\prime},y^{\prime}%
)}{\partial y^{\prime}}$ \ with \ \ $\frac{\partial A_{y}(x^{\prime}%
,y^{\prime})}{\partial x^{\prime}}-B_{z}(x^{\prime},y^{\prime})$, carrying out
the integration with respect to $x^{\prime}$, and by demanding that $f(y)$
\ be independent of $\ x$, we finally obtain the following general solution%

\begin{equation}
\Lambda(x,y)=\Lambda(x_{0},y_{0})+%
%TCIMACRO{\dint \limits_{x_{0}}^{x}}%
%BeginExpansion
{\displaystyle\int\limits_{x_{0}}^{x}}
%EndExpansion
A_{x}(x^{\prime},y)dx^{\prime}+%
%TCIMACRO{\dint \limits_{y_{0}}^{y}}%
%BeginExpansion
{\displaystyle\int\limits_{y_{0}}^{y}}
%EndExpansion
A_{y}(x_{0},y^{\prime})dy^{\prime}+\left\{
%TCIMACRO{\dint \limits_{y_{0}}^{y}}%
%BeginExpansion
{\displaystyle\int\limits_{y_{0}}^{y}}
%EndExpansion
dy^{\prime}%
%TCIMACRO{\dint \limits_{x_{0}}^{x}}%
%BeginExpansion
{\displaystyle\int\limits_{x_{0}}^{x}}
%EndExpansion
dx^{\prime}B_{z}(x^{\prime},y^{\prime})+g(x)\right\}  +f(y_{0})
\label{Lambda(x,y)1}%
\end{equation}

\[
with\text{ \ }g(x)\text{ \ }chosen\ \ so\text{ \ }that\text{ \ }\left\{
%TCIMACRO{\dint \limits_{y_{0}}^{y}}%
%BeginExpansion
{\displaystyle\int\limits_{y_{0}}^{y}}
%EndExpansion
dy^{\prime}%
%TCIMACRO{\dint \limits_{x_{0}}^{x}}%
%BeginExpansion
{\displaystyle\int\limits_{x_{0}}^{x}}
%EndExpansion
dx^{\prime}B_{z}(x^{\prime},y^{\prime})+g(x)\right\}  \boldsymbol{:}\text{ is
}\mathsf{independent\ of\ }\ x,
\]
which is basically the example shown earlier in (\ref{static1}) but with
included multiplicities through the extra constant $f(y_{0})$ (which for
simple-connected space can be set to zero)$\boldsymbol{.}$ The result
(\ref{Lambda(x,y)1}) applies to cases where the particle passes through
\textit{different} magnetic fields (recall that $B_{z}={\huge (}%
\boldsymbol{B}_{2}-\boldsymbol{B}_{1}{\huge )}_{z}$) in spatial regions that
are remote to the observation point $(x,y)$. Alternatively, by following the
reverse route (first integrating the 2nd equation of the basic system
(\ref{usualgradcomps})) we would obtain%

\begin{equation}
\Lambda(x,y)-\Lambda(x,y_{0})=\int_{y_{0}}^{y}A_{y}(x,y^{\prime})dy^{\prime
}+\hat{h}(x) \label{2ndintegration(x,y)}%
\end{equation}
and by then substituting the result to the 1st we would have%

\begin{equation}
A_{x}\left(  x,y\right)  =%
%TCIMACRO{\dint \limits_{y_{0}}^{y}}%
%BeginExpansion
{\displaystyle\int\limits_{y_{0}}^{y}}
%EndExpansion
\frac{\partial A_{y}(x,y^{\prime})}{\partial x}dy^{\prime}+\hat{h}^{\prime
}(x)+\frac{\partial\Lambda(x,y_{0})}{\partial x} \label{Ax}%
\end{equation}
which if integrated would lead to%

\begin{equation}
\hat{h}(x)=\hat{h}(x_{0})-\Lambda(x,y_{0})+\Lambda(x_{0},y_{0})-%
%TCIMACRO{\dint \limits_{x_{0}}^{x}}%
%BeginExpansion
{\displaystyle\int\limits_{x_{0}}^{x}}
%EndExpansion
dx^{\prime}%
%TCIMACRO{\dint \limits_{y_{0}}^{y}}%
%BeginExpansion
{\displaystyle\int\limits_{y_{0}}^{y}}
%EndExpansion
dy^{\prime}\frac{\partial A_{y}(x^{\prime},y^{\prime})}{\partial x^{\prime}%
}+\int_{x_{0}}^{x}A_{x}\left(  x^{\prime},y\right)  dx^{\prime}+h(y)
\label{h(x)}%
\end{equation}
with $h(y)$ to be chosen in such a way that the entire right-hand-side of
(\ref{h(x)}) is only a function of $\ x$ (hence independent of $\ y$).
Finally, by substituting \ \ $\frac{\partial A_{y}(x^{\prime},y^{\prime}%
)}{\partial x^{\prime}}$ \ with \ \ $\frac{\partial A_{x}(x^{\prime}%
,y^{\prime})}{\partial y^{\prime}}+B_{z}(x^{\prime},y^{\prime})$, carrying out
the integration with respect to $y^{\prime}$, and by demanding that $\hat
{h}(x)$ \ be independent of $\ y$, we would finally obtain the following
general solution%

\begin{equation}
\Lambda(x,y)=\Lambda(x_{0},y_{0})+\int_{x_{0}}^{x}A_{x}(x^{\prime}%
,y_{0})dx^{\prime}+\int_{y_{0}}^{y}A_{y}(x,y^{\prime})dy^{\prime}+\left\{
{\Huge -}%
%TCIMACRO{\dint \limits_{x_{0}}^{x}}%
%BeginExpansion
{\displaystyle\int\limits_{x_{0}}^{x}}
%EndExpansion
dx^{\prime}%
%TCIMACRO{\dint \limits_{y_{0}}^{y}}%
%BeginExpansion
{\displaystyle\int\limits_{y_{0}}^{y}}
%EndExpansion
dy^{\prime}B_{z}(x^{\prime},y^{\prime})+h(y)\right\}  +\hat{h}(x_{0})
\label{Lambda(x,y)4}%
\end{equation}

\[
with\text{ \ }h(y)\text{ \ }chosen\text{ \ }so\text{ \ }that\text{ \ }\left\{
{\Huge -}%
%TCIMACRO{\dint \limits_{x_{0}}^{x}}%
%BeginExpansion
{\displaystyle\int\limits_{x_{0}}^{x}}
%EndExpansion
dx^{\prime}%
%TCIMACRO{\dint \limits_{y_{0}}^{y}}%
%BeginExpansion
{\displaystyle\int\limits_{y_{0}}^{y}}
%EndExpansion
dy^{\prime}B_{z}(x^{\prime},y^{\prime})+h(y)\right\}  \boldsymbol{:}\text{ is
}\mathsf{independent\ of\ }\ y,
\]
which is basically the example shown earlier in (\ref{static2}) but with
included multiplicities through the extra constant $\hat{h}(x_{0})$. One can
actually show that the two solutions are equivalent (i.e. (\ref{static1}) and
(\ref{static2}) for a simple-connected space are equal\cite{kyriakos}), a fact
that can be proved in a way similar to the $(x,t)$-cases of Section VIII. (For
the case of multiple-connectivity of the two-dimensional space, a discussion
of the actual values of the multiplicities $f(y_{0})$ and $\hat{h}(x_{0})$ was
given earlier in Section V and will be summarized later in this Section).

As we saw in the examples of Section V, in case of a striped-distribution of
the magnetic field difference $B_{z}$, the functions $g(x)$ and $h(y)$ in
(\ref{Lambda(x,y)1}) and (\ref{Lambda(x,y)4}) (or equivalently in
(\ref{static1}) and (\ref{static2})) have to be chosen in ways that are
compatible with their corresponding constraints (stated after
(\ref{Lambda(x,y)1}) and (\ref{Lambda(x,y)4})) and completely analogous to the
above discussed $(x,t)$-cases$\boldsymbol{;}$ by then taking the
\textit{difference} of (\ref{static1}) and (\ref{static2}) we obtain that the
\textquotedblleft Aharonov-Bohm phase\textquotedblright\ (the one originating
from the \textit{closed} line integral of $A$'s) is exactly cancelled by the
additional nonlocal term of the magnetic fields (that the particle passed
through). As already mentioned earlier, this is reminiscent of the
cancellation of phases (broadly speaking, a cancellation between the
\textquotedblleft Aharonov-Bohm phase\textquotedblright\ and the semiclassical
phase picked up by the trajectories) observed in the early experiments of
Werner \& Brill\cite{WernerBrill} for particles passing through full magnetic
fields, and our method seems to provide a very natural
justification$\boldsymbol{:}$ as our results are completely general (and for
delocalized states in a simple-connected region they basically describe the
single-valuedness of $\Lambda$), they are also valid and applicable to cases
of narrow wavepackets (or states that describe semiclassical motion) that pass
through magnetic fields, which \textit{was} the case of the Werner \& Brill
experiments. (A similar cancellation of an electric Aharonov-Bohm phase also
occurs for particles passing through a static electric field as we saw in
Section VII). We conclude that, for static cases, and when particles pass
through fields, the new nonlocal terms reported in this work lead quite
generally to a cancellation of Aharonov-Bohm phases that had earlier been
sketchily noticed and only at the semiclassical level.

Since we already mentioned that the deep origin of the above cancellations is
the single-valuedness of $\Lambda$ in simple-connected space, we should add
for completeness that the rigorous proof of the uniqueness at each spatial
point (single-valuedness) of $\Lambda$ for completely delocalized states in
simple-connected space can be given in a directly analogous way to the proof
given in Section VIII for the $(x,t)$-case, and is not repeated here. What is
probably more important to point out is that the above cancellations for
semiclassical trajectories (that pass through a magnetic field) can
alternatively be understood as a compatibility between the Aharonov-Bohm
fringe-displacement and the trajectory-deflection due to the Lorentz force
(the semiclassical phase picked up due to the optical path difference of the
two deflected trajectories \textit{exactly cancels} (is \textit{opposite in
sign} from) the Aharonov-Bohm phase picked up by the trajectories due to the
enclosed flux). [We may mention that this is also related to the well-known
overall rigid displacement of the single-slit envelopes of the two-slit
diffraction pattern, displacement that occurs if the wavepackets actually pass
through a field]. These issues are further discussed in the final Section,
where some popular reports in the literature (Feynman\cite{Feynman},
Felsager\cite{Felsager}, Batelaan \& Tonomura\cite{Batelaan}) are given a
minor correction (of a sign). Similarly, and by also including time $t$ (and
by again correcting a sign-error propagating in the standard literature) we
will give an explanation of why certain classical arguments (invoking the past
$t$-dependent history of the experimental set up) seem to work well (in giving
the correct result for a static Aharonov-Bohm phase).

\bigskip Another point of interest concerning the above found nonlocal
contributions of fields is the plausible question of \textit{what shape }the
field distributions must have (or more accurately, their part enclosed inside
the \textit{observation rectangle}) so that the enclosed flux can be decoupled
to a sum of functions of separate variables, in order for the solutions
obtained above to be immediately applicable (i.e. for the functions $g(x)$ and
$h(y)$ to be possible to determine$\boldsymbol{:}$ each of them must then only
\textit{partially} cancel the corresponding $x$ \textit{or} $y$ dependence,
respectively). We already provided an example of such a distribution of a
homogeneous$\ B_{z}$ (the triangular one) in Section V (see the nontrivial
results (\ref{triangular1}) and (\ref{triangular2})). And as mentioned in
Section V, in cases of circularly shaped distributions (where the enclosed
flux may not be decoupled in $x$ and $y$ terms), it is advantageous to solve
the system directly in polar coordinates. By following a similar procedure (of
solving the system of PDEs resulting from (\ref{usualgrad})) in polar
coordinates $(\rho,\varphi)$, namely%

\[
\frac{\partial\Lambda(\rho,\varphi)}{\partial\rho}=A_{\rho}(\rho
,\varphi)\qquad and\qquad\frac{1}{\rho}\frac{\partial\Lambda(\rho,\varphi
)}{\partial\varphi}=A_{\varphi}(\rho,\varphi)
\]
with steps completely analogous to the above, one can obtain the following
analogs of solutions (\ref{Lambda(x,y)1}) and (\ref{Lambda(x,y)4}), namely%

\begin{equation}
\Lambda(\rho,\varphi)=\Lambda(\rho_{0},\varphi_{0})+%
%TCIMACRO{\dint \limits_{\rho_{0}}^{\rho}}%
%BeginExpansion
{\displaystyle\int\limits_{\rho_{0}}^{\rho}}
%EndExpansion
A_{\rho}(\rho^{\prime},\varphi)d\rho^{\prime}+%
%TCIMACRO{\dint \limits_{\varphi_{0}}^{\varphi}}%
%BeginExpansion
{\displaystyle\int\limits_{\varphi_{0}}^{\varphi}}
%EndExpansion
\rho_{0}A_{\varphi}(\rho_{0},\varphi^{\prime})d\varphi^{\prime}+\left\{
%TCIMACRO{\dint \limits_{\varphi_{0}}^{\varphi}}%
%BeginExpansion
{\displaystyle\int\limits_{\varphi_{0}}^{\varphi}}
%EndExpansion
d\varphi^{\prime}%
%TCIMACRO{\dint \limits_{\rho_{0}}^{\rho}}%
%BeginExpansion
{\displaystyle\int\limits_{\rho_{0}}^{\rho}}
%EndExpansion
\rho^{\prime}d\rho^{\prime}B_{z}(\rho^{\prime},\varphi^{\prime})+g(\rho
)\right\}  +f(\varphi_{0}) \label{polar1}%
\end{equation}

\begin{equation}
with\ \ g(\rho)\ \ chosen\text{ \ }so\text{ \ }that\ \ \left\{
%TCIMACRO{\dint \limits_{\varphi_{0}}^{\varphi}}%
%BeginExpansion
{\displaystyle\int\limits_{\varphi_{0}}^{\varphi}}
%EndExpansion
d\varphi^{\prime}%
%TCIMACRO{\dint \limits_{\rho_{0}}^{\rho}}%
%BeginExpansion
{\displaystyle\int\limits_{\rho_{0}}^{\rho}}
%EndExpansion
\rho^{\prime}d\rho^{\prime}B_{z}(\rho^{\prime},\varphi^{\prime})+g(\rho
)\right\}  \boldsymbol{:}\text{ is }\mathsf{independent\ of\ }\ \rho,
\label{polarcond1}%
\end{equation}

and%

\begin{equation}
\Lambda(\rho,\varphi)=\Lambda(\rho_{0},\varphi_{0})+%
%TCIMACRO{\dint \limits_{\rho_{0}}^{\rho}}%
%BeginExpansion
{\displaystyle\int\limits_{\rho_{0}}^{\rho}}
%EndExpansion
A_{\rho}(\rho^{\prime},\varphi_{0})d\rho^{\prime}+%
%TCIMACRO{\dint \limits_{\varphi_{0}}^{\varphi}}%
%BeginExpansion
{\displaystyle\int\limits_{\varphi_{0}}^{\varphi}}
%EndExpansion
\rho A_{\varphi}(\rho,\varphi^{\prime})d\varphi^{\prime}+\left\{  -%
%TCIMACRO{\dint \limits_{\rho_{0}}^{\rho}}%
%BeginExpansion
{\displaystyle\int\limits_{\rho_{0}}^{\rho}}
%EndExpansion
\rho^{\prime}d\rho^{\prime}%
%TCIMACRO{\dint \limits_{\varphi_{0}}^{\varphi}}%
%BeginExpansion
{\displaystyle\int\limits_{\varphi_{0}}^{\varphi}}
%EndExpansion
d\varphi^{\prime}B_{z}(\rho^{\prime},\varphi^{\prime})+h(\varphi)\right\}
+\hat{h}(\rho_{0}) \label{polar2}%
\end{equation}

\begin{equation}
with\ \ h(\varphi)\ \ chosen\text{ \ }so\text{ \ }that\ \ \left\{  -%
%TCIMACRO{\dint \limits_{\rho_{0}}^{\rho}}%
%BeginExpansion
{\displaystyle\int\limits_{\rho_{0}}^{\rho}}
%EndExpansion
\rho^{\prime}d\rho^{\prime}%
%TCIMACRO{\dint \limits_{\varphi_{0}}^{\varphi}}%
%BeginExpansion
{\displaystyle\int\limits_{\varphi_{0}}^{\varphi}}
%EndExpansion
d\varphi^{\prime}\in(\rho^{\prime},\varphi^{\prime})+h(\varphi)\right\}
\boldsymbol{:}\text{ is }\mathsf{independent\ of\ }\ \varphi,
\label{polarcond2}%
\end{equation}
and in these, the proper choices of $g(\rho)$ and $h(\varphi)$ will again be
determined by their corresponding conditions, depending on the actual shape of
the $B_{z}$-distribution and the positioning of initial and final points
$(\rho_{0},\varphi_{0})$ and $(\rho,\varphi).$ [Furthermore, the observation
rectangle has now given its place to a slice of a circular section]. These
matters however deserve further investigation, as an application of the above
theory to specific cases.

Finally, for completeness we summarize our findings on the issue of
multiplicities (the constant last terms of (\ref{Lambda(x,y)1}) and
(\ref{Lambda(x,y)4})) in case of spatial multiple-connectivity (such as the
standard magnetic Aharonov-Bohm case, in which we can take $g(x)=0$
\ \textit{and} \ $h(y)=0$, since the enclosed magnetic flux is independent of
both $x$ and $y$). According to the \textquotedblleft
definitions\textquotedblright\ of these last terms (see
(\ref{2ndintegration(x,y)}) and (\ref{1stintegration(x,y)}) where the
functions $\hat{h}$ and $f$ were first introduced) we have%

\begin{equation}
\hat{h}(x_{0})=\Lambda(x_{0},y)-\Lambda(x_{0},y_{0})-\int_{y_{0}}^{y}%
A_{y}(x_{0},y^{\prime})dy^{\prime} \label{Hx0}%
\end{equation}

\begin{equation}
f(y_{0})=\Lambda(x,y_{0})-\Lambda(x_{0},y_{0})-\int_{x_{0}}^{x}A_{x}%
(x^{\prime},y_{0})dx^{\prime}. \label{fy0}%
\end{equation}

If we insist $(x,y)$ to also lie in a physically inaccessible region, then we
have\ $\hat{h}(x_{0})=-$\ \ $f(y_{0})=$ enclosed magnetic flux\ (which is
already a constant, independent of $x$ and $y$). This is because the limit
$y\rightarrow y_{0}$ (for fixed $x_{0}$) that is needed in (\ref{Hx0}) in
order to find $\hat{h}(x_{0})$, is as if we went around the loop in the
positive sense, landing on the initial point $(x_{0},y_{0})\boldsymbol{;}$
similarly, the limit $x\rightarrow x_{0}$ (for fixed $y_{0}$) that is needed
in (\ref{fy0}) in order to find $f(y_{0})$, is as if we went around the loop
in the negative sense, landing on the initial point $(x_{0},y_{0}).$ Since
$f(y_{0})$ cancels out the {\Huge \ }$%
%TCIMACRO{\dint \limits_{y_{0}}^{y}}%
%BeginExpansion
{\displaystyle\int\limits_{y_{0}}^{y}}
%EndExpansion
dy^{\prime}%
%TCIMACRO{\dint \limits_{x_{0}}^{x}}%
%BeginExpansion
{\displaystyle\int\limits_{x_{0}}^{x}}
%EndExpansion
dx^{\prime}B_{z}(x^{\prime},y^{\prime})$ term, and $\hat{h}(x_{0})$ cancels
out the {\Huge \ -}$%
%TCIMACRO{\dint \limits_{x_{0}}^{x}}%
%BeginExpansion
{\displaystyle\int\limits_{x_{0}}^{x}}
%EndExpansion
dx^{\prime}%
%TCIMACRO{\dint \limits_{y_{0}}^{y}}%
%BeginExpansion
{\displaystyle\int\limits_{y_{0}}^{y}}
%EndExpansion
dy^{\prime}B_{z}(x^{\prime},y^{\prime})$ term, the two solutions are then
reduced to the usual solutions of mere $A$-integrals along the two paths (i.e.
the standard Dirac phase, with no nonlocal contributions).

\section{\bigskip Full $\boldsymbol{(x,y,t)}$-case}

Finally, let us look at the spatially-two-dimensional and time-dependent case.
This combines effects of (perpendicular) magnetic fields (which, if present
only in physically-inaccessible regions, can have Aharonov-Bohm consequences)
with the temporal nonlocalities of electric fields (parallel to the plane). By
working again in Cartesian spatial coordinates, we now have to deal with the
full system of PDEs%

\begin{equation}
\frac{\partial\Lambda(x,y,t)}{\partial x}=A_{x}(x,y,t),\qquad\frac
{\partial\Lambda(x,y,t)}{\partial y}=A_{y}(x,y,t),\qquad-\frac{1}{c}%
\frac{\partial\Lambda(x,y,t)}{\partial t}=\phi\left(  x,y,t\right)  .
\label{FullSystem}%
\end{equation}
This exercise is considerably longer than the previous ones but important to
solve, in order to see in what manner the solutions of this system manage to
\textit{combine} the spatial and temporal nonlocal effects found above. There
are now 3!=6 alternative integration routes to follow for solving this system
(and, in addition to this, the results in intermediate steps tend to
proliferate). Let us here for demonstration show the intermediate steps for
only two routes (that will give us at the end 4 results as we will see),
starting with the \textit{second} of (\ref{FullSystem})$\boldsymbol{:}$ by
integrating it we obtain the expected generalization of
(\ref{2ndintegration(x,y)}), namely%

\begin{equation}
\Lambda(x,y,t)-\Lambda(x,y_{0},t)=\int_{y_{0}}^{y}A_{y}(x,y^{\prime
},t)dy^{\prime}+f(x,t) \label{1stintegrationFull}%
\end{equation}
which if substituted to the first of (\ref{FullSystem}) gives (after
integration over $x^{\prime}$) \ a $t$-generalization of (\ref{h(x)}), namely%

\begin{equation}
f(x,t)=f(x_{0},t)-\Lambda(x,y_{0},t)+\Lambda(x_{0},y_{0},t)-%
%TCIMACRO{\dint \limits_{x_{0}}^{x}}%
%BeginExpansion
{\displaystyle\int\limits_{x_{0}}^{x}}
%EndExpansion
dx^{\prime}%
%TCIMACRO{\dint \limits_{y_{0}}^{y}}%
%BeginExpansion
{\displaystyle\int\limits_{y_{0}}^{y}}
%EndExpansion
dy^{\prime}\frac{\partial A_{y}(x^{\prime},y^{\prime},t)}{\partial x^{\prime}%
}+\int_{x_{0}}^{x}A_{x}\left(  x^{\prime},y,t\right)  dx^{\prime}+G(y,t)
\label{f(x,t)}%
\end{equation}
with $G(y,t)$ to be chosen in such a way that the entire right-hand-side of
(\ref{f(x,t)}) is only a function of $\ x$ and $t$ (hence independent of
$\ y$). Finally, by substituting \ \ $\frac{\partial A_{y}(x^{\prime
},y^{\prime},t)}{\partial x^{\prime}}$ \ with \ \ $\frac{\partial
A_{x}(x^{\prime},y^{\prime},t)}{\partial y^{\prime}}+B_{z}(x^{\prime
},y^{\prime},t)$, carrying out the integration with respect to $y^{\prime}$,
and by demanding that $\ f(x,t)$ \ be independent of $\ y$, we obtain the
following temporal generalization of (\ref{Lambda(x,y)4})%

\[
\Lambda(x,y,t)=\Lambda(x_{0},y_{0},t)+\int_{x_{0}}^{x}A_{x}(x^{\prime}%
,y_{0},t)dx^{\prime}+\int_{y_{0}}^{y}A_{y}(x,y^{\prime},t)dy^{\prime}+
\]

\begin{equation}
+\left\{  {\Huge -}%
%TCIMACRO{\dint \limits_{x_{0}}^{x}}%
%BeginExpansion
{\displaystyle\int\limits_{x_{0}}^{x}}
%EndExpansion
dx^{\prime}%
%TCIMACRO{\dint \limits_{y_{0}}^{y}}%
%BeginExpansion
{\displaystyle\int\limits_{y_{0}}^{y}}
%EndExpansion
dy^{\prime}B_{z}(x^{\prime},y^{\prime},t)+G(y,t)\right\}  +f(x_{0},t)
\label{Lambda(x,y,t)213}%
\end{equation}

\[
with\text{ \ }G(y,t)\text{ \ }such\text{ }that\text{ \ \ }\left\{  {\Huge -}%
%TCIMACRO{\dint \limits_{x_{0}}^{x}}%
%BeginExpansion
{\displaystyle\int\limits_{x_{0}}^{x}}
%EndExpansion
dx^{\prime}%
%TCIMACRO{\dint \limits_{y_{0}}^{y}}%
%BeginExpansion
{\displaystyle\int\limits_{y_{0}}^{y}}
%EndExpansion
dy^{\prime}B_{z}(x^{\prime},y^{\prime},t)+G(y,t)\right\}  \boldsymbol{:}\text{
\ is }\mathsf{independent\ of\ }\ y.
\]

From this point on, the third equation of the system (\ref{FullSystem}) is
getting involved to determine the nontrivial effect of scalar potentials on
$G(y,t)\boldsymbol{;}$ by combining it with (\ref{Lambda(x,y,t)213}) there
results a wealth of patterns$\boldsymbol{:}$ integration with respect to
$t^{\prime}$ leads to%

\[
G(y,t)=G(y,t_{0})-\Lambda(x_{0},y_{0},t)+\Lambda(x_{0},y_{0},t_{0}%
)-f(x_{0},t)+f(x_{0},t_{0})-c%
%TCIMACRO{\dint \limits_{t_{0}}^{t}}%
%BeginExpansion
{\displaystyle\int\limits_{t_{0}}^{t}}
%EndExpansion
\phi(x,y,t^{\prime})dt^{\prime}-
\]

\begin{equation}
-\left[
%TCIMACRO{\dint \limits_{t_{0}}^{t}}%
%BeginExpansion
{\displaystyle\int\limits_{t_{0}}^{t}}
%EndExpansion
dt^{\prime}%
%TCIMACRO{\dint \limits_{x_{0}}^{x}}%
%BeginExpansion
{\displaystyle\int\limits_{x_{0}}^{x}}
%EndExpansion
dx^{\prime}\frac{\partial A_{x}(x^{\prime},y_{0},t^{\prime})}{\partial
t^{\prime}}+%
%TCIMACRO{\dint \limits_{t_{0}}^{t}}%
%BeginExpansion
{\displaystyle\int\limits_{t_{0}}^{t}}
%EndExpansion
dt^{\prime}%
%TCIMACRO{\dint \limits_{y_{0}}^{y}}%
%BeginExpansion
{\displaystyle\int\limits_{y_{0}}^{y}}
%EndExpansion
dy^{\prime}\frac{\partial A_{y}(x,y^{\prime},t^{\prime})}{\partial t^{\prime}%
}\right]  +%
%TCIMACRO{\dint \limits_{t_{0}}^{t}}%
%BeginExpansion
{\displaystyle\int\limits_{t_{0}}^{t}}
%EndExpansion
dt^{\prime}%
%TCIMACRO{\dint \limits_{x_{0}}^{x}}%
%BeginExpansion
{\displaystyle\int\limits_{x_{0}}^{x}}
%EndExpansion
dx^{\prime}%
%TCIMACRO{\dint \limits_{y_{0}}^{y}}%
%BeginExpansion
{\displaystyle\int\limits_{y_{0}}^{y}}
%EndExpansion
dy^{\prime}\frac{\partial B_{z}(x^{\prime},y^{\prime},t^{^{\prime}})}{\partial
t^{\prime}}+F(x,y) \label{med1}%
\end{equation}
with $F(x,y)$ to be chosed in such a way that the entire right-hand-side of
(\ref{med1}) is only a function of $(y,t),$ hence independent of $x.$ In
(\ref{med1}) there are two possible ways to determine the term in brackets,
and another two ways to determine the term containing $B_{z}.$ The easiest to
follow (the one that more directly leads to the final \textit{conditions} that
the functions $F(x,y)$ and $G(y,t_{0})$ are required to satisfy)
is$\boldsymbol{:}$ (i) to substitute $\frac{\partial A_{x}(x^{\prime}%
,y_{0},t^{\prime})}{\partial t^{\prime}}$ \ with \ \ $-c\left(  E_{x}%
(x^{\prime},y_{0},t^{\prime})+\frac{\partial\phi(x^{\prime},y_{0},t^{\prime}%
)}{\partial x^{\prime}}\right)  $ (and similarly for $\frac{\partial
A_{y}(x,y^{\prime},t^{\prime})}{\partial t^{\prime}}$), and \ (ii) to use the
proviso that magnetic and electric fields are connected through the Faraday's
law of Induction, namely \ $\frac{\partial B_{z}(x^{\prime},y^{\prime
},t^{\prime})}{\partial t^{\prime}}=-c\left(  \frac{\partial E_{y}(x^{\prime
},y^{\prime},t^{\prime})}{\partial x^{\prime}}-\frac{\partial E_{x}(x^{\prime
},y^{\prime},t^{\prime})}{\partial y^{\prime}}\right)  .$ These substitutions
lead to cancellations of several intermediate quantities in
(\ref{Lambda(x,y,t)213}) and (\ref{med1}) and lead to the final result%

\[
\Lambda(x,y,t)=\Lambda(x_{0},y_{0},t_{0})+\int_{x_{0}}^{x}A_{x}(x^{\prime
},y_{0},t)dx^{\prime}+\int_{y_{0}}^{y}A_{y}(x,y^{\prime},t)dy^{\prime}-%
%TCIMACRO{\dint \limits_{x_{0}}^{x}}%
%BeginExpansion
{\displaystyle\int\limits_{x_{0}}^{x}}
%EndExpansion
dx^{\prime}%
%TCIMACRO{\dint \limits_{y_{0}}^{y}}%
%BeginExpansion
{\displaystyle\int\limits_{y_{0}}^{y}}
%EndExpansion
dy^{\prime}B_{z}(x^{\prime},y^{\prime},t)+G(y,t_{0})-
\]

\begin{equation}
-c%
%TCIMACRO{\dint \limits_{t_{0}}^{t}}%
%BeginExpansion
{\displaystyle\int\limits_{t_{0}}^{t}}
%EndExpansion
\phi(x_{0},y_{0},t^{\prime})dt^{\prime}+c%
%TCIMACRO{\dint \limits_{t_{0}}^{t}}%
%BeginExpansion
{\displaystyle\int\limits_{t_{0}}^{t}}
%EndExpansion
dt^{\prime}%
%TCIMACRO{\dint \limits_{x_{0}}^{x}}%
%BeginExpansion
{\displaystyle\int\limits_{x_{0}}^{x}}
%EndExpansion
dx^{\prime}E_{x}(x^{\prime},y,t^{\prime})+c%
%TCIMACRO{\dint \limits_{t_{0}}^{t}}%
%BeginExpansion
{\displaystyle\int\limits_{t_{0}}^{t}}
%EndExpansion
dt^{\prime}%
%TCIMACRO{\dint \limits_{y_{0}}^{y}}%
%BeginExpansion
{\displaystyle\int\limits_{y_{0}}^{y}}
%EndExpansion
dy^{\prime}E_{y}(x_{0},y^{\prime},t^{\prime})+F(x,y)+f(x_{0},t_{0})
\label{LambdaFull1}%
\end{equation}
with the functions $G(y,t_{0})$ \ and $\ F(x,y)$ \ to be chosen in such a way
as to satisfy the following 3 independent conditions$\boldsymbol{:}$%

\begin{equation}
\left\{  G(y,t_{0})-%
%TCIMACRO{\dint \limits_{x_{0}}^{x}}%
%BeginExpansion
{\displaystyle\int\limits_{x_{0}}^{x}}
%EndExpansion
dx^{\prime}%
%TCIMACRO{\dint \limits_{y_{0}}^{y}}%
%BeginExpansion
{\displaystyle\int\limits_{y_{0}}^{y}}
%EndExpansion
dy^{\prime}B_{z}(x^{\prime},y^{\prime},t_{0})\right\}  \boldsymbol{:}%
\ is\text{ \ }\mathsf{independent\ of\ }\ y, \label{Gcondition}%
\end{equation}
which is of course a special case of the condition on $G(y,t)$ above (see
after (\ref{Lambda(x,y,t)213})), and the other 2 turn out to be of the form%

\begin{equation}
\left\{  F(x,y)+c%
%TCIMACRO{\dint \limits_{t_{0}}^{t}}%
%BeginExpansion
{\displaystyle\int\limits_{t_{0}}^{t}}
%EndExpansion
dt^{\prime}%
%TCIMACRO{\dint \limits_{x_{0}}^{x}}%
%BeginExpansion
{\displaystyle\int\limits_{x_{0}}^{x}}
%EndExpansion
dx^{\prime}E_{x}(x^{\prime},y,t^{\prime})\right\}  \boldsymbol{:}\ is\text{
\ }\mathsf{independent\ of\ }\ x, \label{F(x,y)condition1}%
\end{equation}

\begin{equation}
\left\{  F(x,y)+c%
%TCIMACRO{\dint \limits_{t_{0}}^{t}}%
%BeginExpansion
{\displaystyle\int\limits_{t_{0}}^{t}}
%EndExpansion
dt^{\prime}%
%TCIMACRO{\dint \limits_{y_{0}}^{y}}%
%BeginExpansion
{\displaystyle\int\limits_{y_{0}}^{y}}
%EndExpansion
dy^{\prime}E_{y}(x,y^{\prime},t^{\prime})\right\}  \boldsymbol{:}\ is\text{
\ }\mathsf{independent\ of\ }\ y. \label{F(x,y)condition2}%
\end{equation}
It should be noted (for the reader who wants to follow all the steps) that the
final condition (\ref{F(x,y)condition2}) does \textit{not} come out
\textit{directly} as the other two$\boldsymbol{;}$ because the function
$G(y,t)$ has disappeared from the final form (\ref{LambdaFull1}), one needs to
\textit{separately} impose the condition above for $G(y,t)$ \ {\Huge (}namely
$\left\{  {\Huge -}%
%TCIMACRO{\dint \limits_{x_{0}}^{x}}%
%BeginExpansion
{\displaystyle\int\limits_{x_{0}}^{x}}
%EndExpansion
dx^{\prime}%
%TCIMACRO{\dint \limits_{y_{0}}^{y}}%
%BeginExpansion
{\displaystyle\int\limits_{y_{0}}^{y}}
%EndExpansion
dy^{\prime}B_{z}(x^{\prime},y^{\prime},t)+G(y,t)\right\}  \boldsymbol{:}$
$\mathsf{independent\ of\ }\ y${\Huge )} directly on the form (\ref{med1}%
)$\boldsymbol{;}$ and in so doing, it is advantageous to interchange
integrations (namely, do the $t^{\prime}$-integral first) \ in the $B_{z}%
$-term of (\ref{med1}), so that $%
%TCIMACRO{\dint \limits_{t_{0}}^{t}}%
%BeginExpansion
{\displaystyle\int\limits_{t_{0}}^{t}}
%EndExpansion
dt^{\prime}%
%TCIMACRO{\dint \limits_{x_{0}}^{x}}%
%BeginExpansion
{\displaystyle\int\limits_{x_{0}}^{x}}
%EndExpansion
dx^{\prime}%
%TCIMACRO{\dint \limits_{y_{0}}^{y}}%
%BeginExpansion
{\displaystyle\int\limits_{y_{0}}^{y}}
%EndExpansion
dy^{\prime}\frac{\partial B_{z}(x^{\prime},y^{\prime},t^{^{\prime}})}{\partial
t^{\prime}}=%
%TCIMACRO{\dint \limits_{x_{0}}^{x}}%
%BeginExpansion
{\displaystyle\int\limits_{x_{0}}^{x}}
%EndExpansion
dx^{\prime}%
%TCIMACRO{\dint \limits_{y_{0}}^{y}}%
%BeginExpansion
{\displaystyle\int\limits_{y_{0}}^{y}}
%EndExpansion
dy^{\prime}{\huge (}B_{z}(x^{\prime},y^{\prime},t)-B_{z}(x^{\prime},y^{\prime
},t_{0}){\huge )}$, and then impose the (less stringent) condition
(\ref{Gcondition}) on $G(y,t_{0})\boldsymbol{;}$ by following this strategy,
after a number of cancellations of intermediate quantities one finally obtains
the 3rd condition (\ref{F(x,y)condition2}) on $F(x,y).$ (As for the constant
quantity\ $f(x_{0},t_{0})$ appearing in (\ref{LambdaFull1}), this again
describes possible effects of multiple-connectivity at the instant $t_{0}$
(which are absent for simple-connected spacetimes, but will be crucial in the
discussion of the van Kampen thought-experiment to be discussed later)).

\bigskip Eqn. (\ref{LambdaFull1}) was our first solution. It is now crucial to
note that an alternative form of solution (with the functions $G^{\prime}s$
and $F$ satisfying the \textit{same} conditions as above) can be derived if,
in the term in brackets of (\ref{med1}) we merely interchange integrations,
leaving therefore $A$'s everywhere rather than introducing electric
fields$\boldsymbol{;}$ following at the same time the above strategy of
changing the ordering of integrations in the $B_{z}$-term as well (without
therefore using Faraday's law) this alternative form of solution turns out to be%

\[
\Lambda(x,y,t)=\Lambda(x_{0},y_{0},t_{0})+\int_{x_{0}}^{x}A_{x}(x^{\prime
},y_{0},t)dx^{\prime}+\int_{y_{0}}^{y}A_{y}(x,y^{\prime},t)dy^{\prime}-%
%TCIMACRO{\dint \limits_{x_{0}}^{x}}%
%BeginExpansion
{\displaystyle\int\limits_{x_{0}}^{x}}
%EndExpansion
dx^{\prime}%
%TCIMACRO{\dint \limits_{y_{0}}^{y}}%
%BeginExpansion
{\displaystyle\int\limits_{y_{0}}^{y}}
%EndExpansion
dy^{\prime}B_{z}(x^{\prime},y^{\prime},t_{0})+G(y,t_{0})-
\]

\begin{equation}
-c%
%TCIMACRO{\dint \limits_{t_{0}}^{t}}%
%BeginExpansion
{\displaystyle\int\limits_{t_{0}}^{t}}
%EndExpansion
\phi(x_{0},y_{0},t^{\prime})dt^{\prime}+c%
%TCIMACRO{\dint \limits_{t_{0}}^{t}}%
%BeginExpansion
{\displaystyle\int\limits_{t_{0}}^{t}}
%EndExpansion
dt^{\prime}%
%TCIMACRO{\dint \limits_{x_{0}}^{x}}%
%BeginExpansion
{\displaystyle\int\limits_{x_{0}}^{x}}
%EndExpansion
dx^{\prime}E_{x}(x^{\prime},y_{0},t^{\prime})+c%
%TCIMACRO{\dint \limits_{t_{0}}^{t}}%
%BeginExpansion
{\displaystyle\int\limits_{t_{0}}^{t}}
%EndExpansion
dt^{\prime}%
%TCIMACRO{\dint \limits_{y_{0}}^{y}}%
%BeginExpansion
{\displaystyle\int\limits_{y_{0}}^{y}}
%EndExpansion
dy^{\prime}E_{y}(x,y^{\prime},t^{\prime})+F(x,y)+f(x_{0},t_{0}).
\label{LambdaFull2}%
\end{equation}
In this alternative solution we note that, in comparison with
(\ref{LambdaFull1}), the line-integrals of $\ \boldsymbol{E}$ \ have changed
to the \textit{other} alternative \textquotedblleft path\textquotedblright%
\ (note the difference in the placement of the coordinates of the initial
point $(x_{0},y_{0})$ in the arguments of $E_{x}$ and $E_{y}$) and they happen
to have the same sense as the $\boldsymbol{A}$-integrals, while simultaneously
the magnetic flux difference shows up with its value at the initial time
$t_{0}$ rather than at $t$. This alternative form will be shown to be useful
in cases where we want to directly compare physical situations in the present
(at time $t$) and in the past (at time $t_{0}$), and the above noted change of
sense of $\boldsymbol{E}$-integrals (compared to (\ref{LambdaFull1})) will be
crucial in the discussion that follows (in Section XI).

Once again the reader can directly verify that (\ref{LambdaFull1}) or
(\ref{LambdaFull2}) indeed satisfy the basic input system (\ref{FullSystem}).
(This verification is a bit more tedious than the earlier ones but
straightforward, and is not shown here).

\bigskip But in order to discuss the van Kampen case, namely an enclosed (and
physically inaccessible) magnetic flux (which however is
\textit{time-dependent})\textit{, }it is important to have the analogous forms
through a reverse route, namely starting with (integrating) the \textit{first}
of (\ref{FullSystem}) and then substituting the result to the
second$\boldsymbol{;}$ in this way we will at the end have the reverse
\textquotedblleft path\textquotedblright\ of $\boldsymbol{A}$-integrals, so
that by taking the \textit{difference} of the resulting solution and the above
solution (\ref{LambdaFull1}) (or (\ref{LambdaFull2})) will lead to the
\textit{closed} line integral of $\boldsymbol{A}$\textbf{ }which will be
immediately related to the van Kampen's magnetic flux (at the instant $t$). By
following then this route, and by applying a similar strategy at every
intermediate step, we finally obtain the following solution (the spatially
\textquotedblleft dual\textquotedblright\ to (\ref{LambdaFull1})), namely%

\[
\Lambda(x,y,t)=\Lambda(x_{0},y_{0},t_{0})+\int_{x_{0}}^{x}A_{x}(x^{\prime
},y,t)dx^{\prime}+\int_{y_{0}}^{y}A_{y}(x_{0},y^{\prime},t)dy^{\prime}+%
%TCIMACRO{\dint \limits_{x_{0}}^{x}}%
%BeginExpansion
{\displaystyle\int\limits_{x_{0}}^{x}}
%EndExpansion
dx^{\prime}%
%TCIMACRO{\dint \limits_{y_{0}}^{y}}%
%BeginExpansion
{\displaystyle\int\limits_{y_{0}}^{y}}
%EndExpansion
dy^{\prime}B_{z}(x^{\prime},y^{\prime},t)+\hat{G}(x,t_{0})-
\]

\begin{equation}
-c%
%TCIMACRO{\dint \limits_{t_{0}}^{t}}%
%BeginExpansion
{\displaystyle\int\limits_{t_{0}}^{t}}
%EndExpansion
\phi(x_{0},y_{0},t^{\prime})dt^{\prime}+c%
%TCIMACRO{\dint \limits_{t_{0}}^{t}}%
%BeginExpansion
{\displaystyle\int\limits_{t_{0}}^{t}}
%EndExpansion
dt^{\prime}%
%TCIMACRO{\dint \limits_{x_{0}}^{x}}%
%BeginExpansion
{\displaystyle\int\limits_{x_{0}}^{x}}
%EndExpansion
dx^{\prime}E_{x}(x^{\prime},y_{0},t^{\prime})+c%
%TCIMACRO{\dint \limits_{t_{0}}^{t}}%
%BeginExpansion
{\displaystyle\int\limits_{t_{0}}^{t}}
%EndExpansion
dt^{\prime}%
%TCIMACRO{\dint \limits_{y_{0}}^{y}}%
%BeginExpansion
{\displaystyle\int\limits_{y_{0}}^{y}}
%EndExpansion
dy^{\prime}E_{y}(x,y^{\prime},t^{\prime})+F(x,y)+\hat{h}(y_{0},t_{0})
\label{LambdaFull4}%
\end{equation}
with the functions $\hat{G}(x,t_{0})$ \ and $\ F(x,y)$ \ to be chosen in such
a way as to satisfy the following 3 independent conditions$\boldsymbol{:}$%

\begin{equation}
\left\{  \hat{G}(x,t_{0})+%
%TCIMACRO{\dint \limits_{y_{0}}^{y}}%
%BeginExpansion
{\displaystyle\int\limits_{y_{0}}^{y}}
%EndExpansion
dy^{\prime}%
%TCIMACRO{\dint \limits_{x_{0}}^{x}}%
%BeginExpansion
{\displaystyle\int\limits_{x_{0}}^{x}}
%EndExpansion
dx^{\prime}B_{z}(x^{\prime},y^{\prime},t_{0})\right\}  \boldsymbol{:}%
\ is\ \ \mathsf{independent\ of\ }\ x, \label{Gcondition3}%
\end{equation}

\bigskip%
\begin{equation}
\left\{  F(x,y)+c%
%TCIMACRO{\dint \limits_{t_{0}}^{t}}%
%BeginExpansion
{\displaystyle\int\limits_{t_{0}}^{t}}
%EndExpansion
dt^{\prime}%
%TCIMACRO{\dint \limits_{x_{0}}^{x}}%
%BeginExpansion
{\displaystyle\int\limits_{x_{0}}^{x}}
%EndExpansion
dx^{\prime}E_{x}(x^{\prime},y,t^{\prime})\right\}  \boldsymbol{:}%
\ is\ \ \mathsf{independent\ of\ }\ x, \label{F(x,y)condition3}%
\end{equation}

\bigskip%
\begin{equation}
\left\{  F(x,y)+c%
%TCIMACRO{\dint \limits_{t_{0}}^{t}}%
%BeginExpansion
{\displaystyle\int\limits_{t_{0}}^{t}}
%EndExpansion
dt^{\prime}%
%TCIMACRO{\dint \limits_{y_{0}}^{y}}%
%BeginExpansion
{\displaystyle\int\limits_{y_{0}}^{y}}
%EndExpansion
dy^{\prime}E_{y}(x,y^{\prime},t^{\prime})\right\}  \boldsymbol{:}%
\ is\ \ \mathsf{independent\ of\ }\ y, \label{F(x,y)condition4}%
\end{equation}
where again for the above results the Faraday's law was crucial. The
corresponding analog of the alternative form (\ref{LambdaFull2}) (where
$B_{z}$ appears at $t_{0}$) is more important and turns out to be%

\[
\Lambda(x,y,t)=\Lambda(x_{0},y_{0},t_{0})+\int_{x_{0}}^{x}A_{x}(x^{\prime
},y,t)dx^{\prime}+\int_{y_{0}}^{y}A_{y}(x_{0},y^{\prime},t)dy^{\prime}+%
%TCIMACRO{\dint \limits_{x_{0}}^{x}}%
%BeginExpansion
{\displaystyle\int\limits_{x_{0}}^{x}}
%EndExpansion
dx^{\prime}%
%TCIMACRO{\dint \limits_{y_{0}}^{y}}%
%BeginExpansion
{\displaystyle\int\limits_{y_{0}}^{y}}
%EndExpansion
dy^{\prime}B_{z}(x^{\prime},y^{\prime},t_{0})+\hat{G}(x,t_{0})-
\]

\begin{equation}
-c%
%TCIMACRO{\dint \limits_{t_{0}}^{t}}%
%BeginExpansion
{\displaystyle\int\limits_{t_{0}}^{t}}
%EndExpansion
\phi(x_{0},y_{0},t^{\prime})dt^{\prime}+c%
%TCIMACRO{\dint \limits_{t_{0}}^{t}}%
%BeginExpansion
{\displaystyle\int\limits_{t_{0}}^{t}}
%EndExpansion
dt^{\prime}%
%TCIMACRO{\dint \limits_{x_{0}}^{x}}%
%BeginExpansion
{\displaystyle\int\limits_{x_{0}}^{x}}
%EndExpansion
dx^{\prime}E_{x}(x^{\prime},y,t^{\prime})+c%
%TCIMACRO{\dint \limits_{t_{0}}^{t}}%
%BeginExpansion
{\displaystyle\int\limits_{t_{0}}^{t}}
%EndExpansion
dt^{\prime}%
%TCIMACRO{\dint \limits_{y_{0}}^{y}}%
%BeginExpansion
{\displaystyle\int\limits_{y_{0}}^{y}}
%EndExpansion
dy^{\prime}E_{y}(x_{0},y^{\prime},t^{\prime})+F(x,y)+\hat{h}(y_{0},t_{0})
\label{LambdaFIN}%
\end{equation}
with $\hat{G}(x,t_{0})$ and $F(x,y)$ following the same 3 conditions above.
The constant term $\hat{h}(y_{0},t_{0})$ again describes possible
multiplicities at the instant $t_{0}\boldsymbol{;}$ it is absent for
simple-connected spacetimes, but will be crucial in the discussion of the van
Kampen thought-experiment.

In (\ref{LambdaFull4}) (and in (\ref{LambdaFIN})), note the \textquotedblleft
alternative paths\textquotedblright\ (compared to solution (\ref{LambdaFull1})
(and (\ref{LambdaFull2}))) of line integrals of $\boldsymbol{A}$'s (or of
$\boldsymbol{E}$'s). But the most crucial element for what follows is the use
of forms (\ref{LambdaFull2}) and (\ref{LambdaFIN}) (where $B_{z}$ only appears
at $t_{0}$), and the fact that, within each solution, the sense of
$\boldsymbol{A}$-integrals is the \textit{same} as the sense of the
$\boldsymbol{E}$-integrals. (This is \textit{not} true in the other solutions
where $B_{z}(..,t)$ appears). These facts will be crucial to the discussion
that follows, which briefly addresses the so called van Kampen
\textquotedblleft paradox\textquotedblright.

\section{\bigskip The van Kampen thought-experiment -- Causal Issues hidden in
the above solutions}

In that early work\cite{vanKampen} van Kampen considered a genuine
Aharonov-Bohm case, with a magnetic flux (physically inaccessible to the
particle) which, however, is time-dependent$\boldsymbol{:}$ van Kampen
envisaged turning on the flux very late, or equivalently, observing the
interference of the two wavepackets (on a distant screen) very early, earlier
than the time it takes light to travel the distance to the screen, hence using
the (instantaneous nature of the) Aharonov-Bohm phase to transmit information
(on the existence of a confined magnetic flux somewhere in space)
\textit{superluminally}. Indeed, the Aharonov-Bohm phase at any later instant
$t$ is determined by differences of $\frac{q}{\hbar c}\Lambda(\mathbf{r},t)$,
\ with $\ \Lambda(\mathbf{r},t)=\int_{\mathbf{r}_{0}}^{\mathbf{r}%
}\boldsymbol{A}(\mathbf{r}^{\prime},t)\boldsymbol{.}d\mathbf{r}^{\prime}+$
$const.$ (which basically results as a special case (but in higher
dimensionality) of the incorrect expression (\ref{BrownHolland}) in the
temporal gauge $\phi=0$, the constant being $\Lambda(\mathbf{r}_{0},t_{0})$).
However, let us for this case utilize instead our results (\ref{LambdaFull2})
and (\ref{LambdaFIN}) above, where we have the additional appearance of the
nonlocal $E$-terms (and of the $B_{z}$-term at $t_{0}$).

In order to be slightly more general, let us for example assume that the
inaccessible magnetic flux had the value $\Phi(t_{0})$ at $t_{0}$, and then it
started changing with time. By using a narrow wavepacket picture like van
Kampen, we can then subtract (\ref{LambdaFull2}) and (\ref{LambdaFIN}) in
order to find the phase difference at a time $t$ that is smaller than the time
required for light to reach the observation point $(x,y)$ (i.e. $t<$ $\frac
{L}{c}$, with $L$ the corresponding distance). For a spatially-confined
magnetic flux $\Phi(t)$, the functions $G,$ $\hat{G}$ and $F$ in the above
solutions can all be taken zero$\boldsymbol{:}$ their conditions are all
satisfied for a flux $\Phi(t)$ that is not spatially-extended (hence, from
(\ref{Gcondition}) and (\ref{Gcondition3}) we obtain $G=\hat{G}=0$) and, for
$t<$ $\frac{L}{c}$, the integrals of $E_{x}$ and $E_{y}$ in conditions
(\ref{F(x,y)condition1}) and (\ref{F(x,y)condition2}) (or in
(\ref{F(x,y)condition3}) and (\ref{F(x,y)condition4})) are already independent
of \textit{both} $x$ and $y$ (since $E_{x}(x,y,t^{\prime})=E_{y}%
(x,y,t^{\prime})=0$ for all $t^{\prime}<t<\frac{L}{c}$, with $(x,y)$ the
observation point on the screen, and therefore all integrations of $E_{x}$ and
$E_{y}$ with respect to $x^{\prime}$ and $y^{\prime}$ will give results that
are \textit{independent of the integration upper limits }$x$\textit{ and
}$y\boldsymbol{;}$ hence $F=0$). Moreover, the multiplicities $(f$ and
$\hat{h})$ \ lead to cancellation of the $B_{z}$-terms (at $t_{0}$) as
outlined in the static case earlier (end of Section IX). By choosing then the
temporal gauge $\phi=0,$ we have for the difference (\ref{LambdaFull2}) $-$
(\ref{LambdaFIN}) at the point and instant of observation the following result%

\[
\Delta\Lambda(x,y,t)=\int_{x_{0}}^{x}A_{x}(x^{\prime},y_{0},t)dx^{\prime}%
+\int_{y_{0}}^{y}A_{y}(x,y^{\prime},t)dy^{\prime}-\int_{x_{0}}^{x}%
A_{x}(x^{\prime},y,t)dx^{\prime}-\int_{y_{0}}^{y}A_{y}(x_{0},y^{\prime
},t)dy^{\prime}+
\]

\begin{equation}
+c%
%TCIMACRO{\dint \limits_{t_{0}}^{t}}%
%BeginExpansion
{\displaystyle\int\limits_{t_{0}}^{t}}
%EndExpansion
dt^{\prime}\left\{
%TCIMACRO{\dint \limits_{x_{0}}^{x}}%
%BeginExpansion
{\displaystyle\int\limits_{x_{0}}^{x}}
%EndExpansion
dx^{\prime}E_{x}(x^{\prime},y_{0},t^{\prime})+%
%TCIMACRO{\dint \limits_{y_{0}}^{y}}%
%BeginExpansion
{\displaystyle\int\limits_{y_{0}}^{y}}
%EndExpansion
dy^{\prime}E_{y}(x,y^{\prime},t^{\prime})-%
%TCIMACRO{\dint \limits_{x_{0}}^{x}}%
%BeginExpansion
{\displaystyle\int\limits_{x_{0}}^{x}}
%EndExpansion
dx^{\prime}E_{x}(x^{\prime},y,t^{\prime})-%
%TCIMACRO{\dint \limits_{y_{0}}^{y}}%
%BeginExpansion
{\displaystyle\int\limits_{y_{0}}^{y}}
%EndExpansion
dy^{\prime}E_{y}(x_{0},y^{\prime},t^{\prime})\right\}  . \label{DeltaLambda}%
\end{equation}
In (\ref{DeltaLambda}) the sum of the four $A$-integrals gives the
\textit{closed} line-integral of vector $\boldsymbol{A}$ around the
\textit{observation rectangle }at time $t$ (in the positive sense) and it is
equal to the instantaneous magnetic flux $\Phi(t)$ (that leads to the
\textquotedblleft usual\textquotedblright\ magnetic Aharonov-Bohm
phase)$\boldsymbol{;}$ the sum of the four $E$-integrals inside the brackets
in the last terms (originating from our nonlocal contributions) gives the
\textit{closed} line-integral of vector $\boldsymbol{E}$ around the same
rectangle at any arbitrary $t^{\prime}$, and in the same (positive) sense
(something we wouldn't have if we had taken the first type of solutions,
(\ref{LambdaFull1}) and (\ref{LambdaFull4}) $-$ this signifying the importance
of taking the right form, the one that contains $B_{z}$ at $t_{0}$ (with the
$t$-propagation of $B_{z}$ \textit{having already been incorporated} in the
$E_{x}$ and $E_{y}$ terms of (\ref{LambdaFull2}) and (\ref{LambdaFIN}))). By
denoting therefore the closed loop integral (around the rectangle) as $%
%TCIMACRO{\doint }%
%BeginExpansion
{\displaystyle\oint}
%EndExpansion
$ always in the positive sense (and with the understanding that the
rectangle's upper right corner is the spatial point of observation $(x,y)$),
(\ref{DeltaLambda}) reads%

\begin{equation}
\Delta\Lambda(x,y,t)=%
%TCIMACRO{\doint }%
%BeginExpansion
{\displaystyle\oint}
%EndExpansion
\boldsymbol{A}(\mathbf{r}^{\prime},t)\boldsymbol{.}d\mathbf{r}^{\prime}+c%
%TCIMACRO{\dint \limits_{t_{0}}^{t}}%
%BeginExpansion
{\displaystyle\int\limits_{t_{0}}^{t}}
%EndExpansion
dt^{\prime}%
%TCIMACRO{\doint }%
%BeginExpansion
{\displaystyle\oint}
%EndExpansion
\boldsymbol{E}(\mathbf{r}^{\prime},t^{\prime})\boldsymbol{.}d\mathbf{r}%
^{\prime} \label{DeltaLambdaBrief}%
\end{equation}
which, with $%
%TCIMACRO{\doint }%
%BeginExpansion
{\displaystyle\oint}
%EndExpansion
\boldsymbol{A}(\mathbf{r}^{\prime},t)\boldsymbol{.}d\mathbf{r}^{\prime}%
=\Phi(t)$ \ the instantaneous enclosed magnetic flux and with the help of
Faraday's law $%
%TCIMACRO{\doint }%
%BeginExpansion
{\displaystyle\oint}
%EndExpansion
\boldsymbol{E}(\mathbf{r}^{\prime},t^{\prime})\boldsymbol{.}d\mathbf{r}%
^{\prime}=-\frac{1}{c}\frac{d\Phi(t^{\prime})}{dt^{\prime}},$ \ gives%

\begin{equation}
\Delta\Lambda(x,y,t)=\Phi(t)-{\huge (}\Phi(t)-\Phi(t_{0}){\huge )}=\Phi
(t_{0}). \label{DeltaLambdaFinal}%
\end{equation}
Although $\Delta\Lambda$ is generally $t$-dependent, we obtain the intuitive
result that, for $t<\frac{L}{c}$ (i.e. if the physical information has not yet
reached the screen), the phase-difference turns out to be $t$-independent, and
leads to the magnetic Aharonov-Bohm\ phase that we \textit{would} observe at
$t_{0}$.

This gives an honest resolution of the \textquotedblleft van Kampen
paradox\textquotedblright\ within a canonical formulation, without using any
vague electric Aharonov-Bohm effect argument (since in the gauge chosen
$(\phi=0)$ there are no scalar potentials\ -- and, most importantly,
\textit{there is no multiple-connectivity in }$(x,t)$\textit{-plane} as in the
electric Aharonov-Bohm case\cite{Iddings}). An additional physical element (in
comparison to van Kampen's electric phase interpretation) is that, for the
above cancellation, it is not only the $E$-fields but also the $t$-propagation
in space of the $B_{z}$-fields (the full \textquotedblleft radiation
field\textquotedblright) that plays a role.

Finally, a number of other forms of solutions can be obtained that result from
different ordering of integrations of the system (\ref{FullSystem}) (a full
list of 12 different (but quite long) results is available, and easily
verifiable that they indeed satisfy the system (\ref{FullSystem})). The reader
can follow the strategies suggested here and derive the forms that are
appropriate to particular physical cases of interest that may be different
from the above magnetic case, some potential candidates being the
\textquotedblleft electric analog\textquotedblright\ of the van Kampen
thought-experiment, or its \textit{bound state analog} in nanorings. For the
latter, and especially for 1-D nanorings (or other nanoscopic devices) driven
by a $t$-dependent magnetic flux, the new nonlocal terms are expected to be of
relevance if they are included in standard treatments\cite{LuanTang}, and the
effects are expected to appear in the PetaHertz range. (Similarly we might
expect a role in cases of quantal astrophysical objects due to the large
distances involved (hence retardation effects being more pronounced)).

For the \textquotedblleft electric analog\textquotedblright\ of the van Kampen
case, we note that, although this has never really been discussed in the
literature, nevertheless, it has been essentially briefly mentioned in
Appendix B of Peshkin\cite{Peshkin} (where the case that \textquotedblleft%
\textit{first} the particle exits the cages, and \textit{only then} we switch
on the outside electric field\textquotedblright\ is made, together with the
comment that the results must be \textquotedblleft consistent with ordinary
ideas about Causality\textquotedblright$\boldsymbol{;}$ Peshkin correctly
states: \textquotedblleft One cannot wait for the electron to pass and only
later switch on the field to cause a physical effect\textquotedblright). As
our new nonlocal terms seem to be especially suited for addressing such
Causality issues, let us slightly expand on this point$\boldsymbol{:}$ in this
most authoritative (and carefully written) review of the Aharonov-Bohm effect
in the literature, Peshkin uses (for the electric effect) a solution-form (his
eqn.(B.5) together with (B.6)) based on (\ref{BrownHolland}), i.e. the
\textquotedblleft standard result\textquotedblright\ (but applied to a
spatially-dependent scalar potential) $-$ but he clearly states that it is an
approximation (and actually later in the review, he states that this form
cannot be a solution for all $t$). Indeed, from the present work we learn that
(B.5) and (B.6) is not the solution when the scalar potential depends on
spatial variables (because the spatial variables inside the potential will
give $-$ through its nonzero gradient $-$ an extra vector potential (that will
result from $\nabla\Lambda$), hence an extra minimal substitution in the
Hamiltonian $H$, violating therefore the mapping between two predescribed
systems that we want to achieve). As we saw in the present work, the correct
solution for all $t$ and in all space consists of additional nonlocal terms of
the appropriate form. If we view the form (B.5) and (B.6) of ref.
\cite{Peshkin} as an \textit{ansatz}, then it is understandable why a
\textit{condition} (Peshkin's eqn.(B.8), and later (B.9)) needs to be
\textit{enforced} on the electric field outside the cages (in order for the
extra (annoying) terms (that show up from expansion of the squared minimal
substitution) to vanish and for (B.5) to be a solution). And then Peshkin
notes that the extra condition cannot always be satisfied $-$ \textit{it must
fail} for some times (hence (B.5) is not really the solution for all times),
drawing from this a correct conclusion, namely that \textquotedblleft the
electron must traverse some region where the electric field \textit{has
been}\textquotedblright\ (earlier). However, the causal issue pointed out
above, although mentioned in words, is not dealt with quantitatively. From our
present work, it turns out that the total \textquotedblleft radiation
field\textquotedblright\ outside the cages is crucial in recovering Causality,
in a similar way as in the case presented above in this Section for the usual
(magnetic) version of the van Kampen experiment. In this \textquotedblleft
electric analog\textquotedblright\ that we are discussing now, the
causally-offending part of the electric Aharonov-Bohm phase difference will be
cancelled by a magnetic type of phase, that originates from the magnetic field
that is associated with the $t$-dependence of the electric field
$\boldsymbol{E}$ outside the cages.

It should be reemphasized that the correct quantitative physical behavior for
the above system for all times comes out from the treatment shown in detail in
the present work, with no enforced constraints, but with conditions that come
out naturally from the solution of the PDEs. The results that are derived from
this careful procedure give the full solutions (correct for all space and for
all $t$)$\boldsymbol{:}$ Peshkin's ansatz (B.6) turns out (from an honest and
careful solution of the full PDEs) to be augmented by nonlocal (in time) terms
of the electric fields, and these directly influence the phases of
wavefunctions (by always respecting Causality, with no need of enforced
statements) -- and can even include the contributions of vector potentials and
magnetic fields (through nonlocal magnetic terms in space) associated with the
$t$-variation of the electric field outside the cages, that Peshkin has
omitted. As already mentioned, the total \textquotedblleft radiation
field\textquotedblright\ outside the cages is crucial in recovering Causality,
in a way similar to what was presented in this work for the usual (magnetic)
version of the van Kampen experiment. We conclude that our (exact) results
accomplish precisely what Peshkin has in mind in his discussion (on
Causality), but in a direct and fully quantitative manner, and with \textit{no
ansatz} based on an incorrect form.

\section{Discussion}

\bigskip Trying to evaluate in a broader sense the crucial nonlocal influences
found in all the above physical examples, we should probably reemphasize that
at the level of the basic Lagrangian $L(\mathbf{r},\mathbf{v},t)=\frac{1}%
{2}m\mathbf{v}^{2}+\frac{q}{c}\mathbf{v}\boldsymbol{.}\boldsymbol{A}%
(\mathbf{r},t)-q\phi(\mathbf{r},t)$ \ there are no fields present, and the
view holds in the literature\cite{BrownHome} that electric or magnetic fields
cannot contribute \textit{directly} to the phase. This view originates from
the path-integral treatments widely used (where the Lagrangian determines
directly the phases of Propagators), but, nevertheless, our canonical
formulation treatment shows that fields \textit{do} contribute nonlocally, and
they are actually crucial in recovering Relativistic Causality. Moreover,
path-integral discussions\cite{Troudet} of the van Kampen case use wave
(retarded)-solutions for the vector potentials $\boldsymbol{A}$ (hence they
are treated in Lorenz gauge, which is not sufficiently general$\boldsymbol{:}$
even if $\boldsymbol{A}$ has not yet reached the screen, we can always add a
constant $\boldsymbol{A}$ (a pure gauge) over all space, and there are no more
retarded wave-solutions for the potentials, the path-integral resolution of
the paradox being, therefore, at least incomplete). Our results are
gauge-invariant and take advantage of only the retardation of \textit{fields}
$\boldsymbol{E}$ and $\boldsymbol{B}$\textbf{ }(true in \textit{any} gauge),
and \textit{not }of potentials. In addition, Troudet\cite{Troudet} clearly
(and correctly) states that his treatment is good for not highly-delocalized
states in space, and that in case of delocalization the proper treatment
\textquotedblleft would be much more complicated, and would require a much
more complete analysis\textquotedblright. We believe we have provided one in
this paper. It should be added that in a recent Compendium of Quantum
Physics\cite{Compendium}, the \textquotedblleft van Kampen
paradox\textquotedblright\ still seems to be thought of as remarkable. We
believe that this work has provided a natural and general resolution, and most
importantly, through nonlocal and Relativistically causal propagation of
wavefunction phases (this point being expanded further at the end of the paper).

At several places in this article we have pointed out a number of
\textquotedblleft misconceptions\textquotedblright\ in the literature (mostly
on the uncritical use of the (standard) Dirac phases even for $t$-dependent
vector potentials and spatially-dependent scalar potentials, which is plainly
incorrect for uncorrelated variables), and we have explicitly provided their
\textquotedblleft healing\textquotedblright\ through appropriate nonlocal
field-terms. It should however be emphasized here that this is not a merely
marginal misconception, but it appears all over the place in the literature
(due to the Feynman path-integral bias)$\boldsymbol{;}$ it is even stated by
Feynman himself in volume II of his \textit{Lectures on Physics}%
\cite{Feynman}, namely\textit{\ }that the simple phase factor $\int^{x}A\cdot
d\mathbf{r}^{\prime}-c\int^{t}\phi dt^{\prime}$ is valid generally, i.e. even
for $t$-dependent fields. Similarly, this erroneous generalization is also
explicitly stated in the review on Aharonov-Bohm effects of
Erlichson\cite{Erlichson} that has given a very balanced view of earlier
controversy, and elsewhere $-$ the books of Silverman\cite{Silverman} being
the clearest case that we are aware of with a careful wording about
(\ref{BrownHolland}) being only restrictedly valid (for $t$-independent
$\boldsymbol{A}$'s and $\mathbf{r}$-independent $\phi$'s) $-$ although even
there the nonlocal terms have been missed. We believe that the above
misconceptions (and the overlooking of the nonlocal terms) are the basic
reason why \textquotedblleft it appears that no exact theoretical treatment
has been given\textquotedblright\ (for the electric Aharonov-Bohm effect), as
correctly stated by Peshkin in his Appendix B of Ref. \cite{Peshkin}.

And let us now come to a second type of misconception, that is probably less
important since it has appeared only in semiclassical conditions $-$ but is
essential to mention here, as it also exhibits the merits of our approach (and
the deeper physical understanding that our results can lead to). What we learn
from the generalized Werner \& Brill cancellations pointed out rather
emphatically in this work is that, at the point of observation, the nonlocal
terms of classical remote fields have the tendency to contribute a phase
\textit{of opposite sign} to the \textquotedblleft Aharonov-Bohm
phase\textquotedblright\ (of potentials). We want to point out to the reader
that, for semiclassical trajectories, this is actually descriptive of the
compatibility (or consistency) of the Aharonov-Bohm fringe-displacement and
the associated trajectory-deflection due to the classical forces. Let us for
example look at Fig.15-8 of Feynman\cite{Feynman2}, or at Fig.2.16 of the book
of Felsager\cite{Felsager}, where, classical trajectories are deflected after
they pass through a strip of a homogeneous magnetic field that is placed on
the right of a standard double-slit experimental apparatus. Both authors
determine the semiclassical phase picked up by the trajectories (that have
been deflected by the Lorentz force) and they find that they are consistent
with the Aharonov-Bohm phase (picked up due to the flux enclosed by the same
trajectories). However, it is rather straightforward for the reader to see
that the two phases \textit{have opposite sign }(they are not equal as implied
by the authors). (The reader is also invited to carry out a similar exercise,
with particles passing through an analogous homogeneous electric field on the
right of the double-slit apparatus, that is switched on for a finite duration
$T$, where again the semiclassical phase picked up turns out to be opposite to
the electric Aharonov-Bohm type of phase). Similarly, in the very recent
review of Batelaan \& Tonomura\cite{Batelaan}, their Fig.2 contains visual
information that is very relevant to our discussion: it is a quite descriptive
picture of the wavefronts associated to the classical trajectories, where the
authors state that \textquotedblleft the phase shift calculated in terms of
the Lorentz force is the same as that predicted by the Aharonov-Bohm effect in
terms of the vector potential $A$ circling the magnetic bar\textquotedblright.
The reader, however, should notice once more that the sign of the classical
phase-difference is really opposite to the sign of the Aharonov-Bohm phase.
The phases are not equal as stated, but opposite. All the above examples are
we believe a manifestation \ of the cancellations that have been derived in
the present work (for general quantum states), but here they are just special
cases for semiclassical trajectories. (We could also add that these
cancellations also have to do with the well-known rigid displacement of the
\textquotedblleft single-slit envelope\textquotedblright\ of the two-slit
diffraction pattern in a double-slit experiment with an additional strip of a
magnetic field placed on the right of the apparatus).

In a slightly different vein, we should also point out that the above
cancellations give a justification of why certain semiclassical arguments that
focus on the history of the experimental set up (usually based on Faraday's
law for a $t$-dependent magnetic flux) seem to give at the end a result that
is consistent with the result of a static Aharonov-Bohm arrangement. However,
there is a again an opposite sign that seems to have been largely unnoticed in
such arguments as well (i.e. see the simplest possible argument in
Silverman\cite{Silverman2}, where in his eqn.(1.34) there should be an extra
minus sign). Our above observation essentially describes the fact that,
\textit{if we had actually used} a $t$-dependent magnetic flux, then the
induced electric field (viewed now as a nonlocal term of the present work)
would have cancelled the static Aharonov-Bohm phase. Of course now, this
$t$-dependent experimental set up has not been used (the flux is static) and
we obtain the usual magnetic Aharonov-Bohm phase, but the above argument (of a
\textquotedblleft potential experiment\textquotedblright\ that \textit{could
have been carried out}) takes the \textquotedblleft mystery\textquotedblright%
\ away of why such arguments generally work $-$ although they have to be
corrected with a sign.

Finally, coming back to a broader significance of the new solutions, one may
wonder about possible consequences of the nonlocal terms if these are included
in more general physical models that have a gauge structure (in Condensed
Matter or High Energy Physics). It is also worth mentioning that by following
the same \textquotedblleft unconventional\textquotedblright\ method (of
solution of PDEs) but now applied to the Maxwell's equations for the electric
and magnetic fields, we obtained the corresponding nonlocal terms, and we
found that these essentially demonstrate the causal propagation of the
radiation electric and magnetic fields outside physically inaccessible
confined sources (i.e. solenoids or electric cages). Although this is of
course widely known at the level of classical fields, a major conclusion that
can be drawn from the present work (at the level of gauge transformations) is
that, a corresponding Causality may exist at the level of quantum mechanical
phases as well, and this is enforced by the nonlocal terms in $t$-dependent
cases. It strongly indicates that the nonlocal terms found here at the level
of quantum mechanical phases reflect a causal propagation of wavefunction
phases \textbf{in the Schr\"{o}dinger picture} (at least one part of them, the
one containing the fields, which competes with the Aharonov-Bohm types of
phases containing the potentials). This is an entirely new concept (given the
local nature but also the nonrelativistic character of the Schr\"{o}dinger
equation) and deserves to be further explored. It would indeed be worth
investigating possible applications of the above results (of nonlocal phases
of wavefunctions, solutions of the local Schr\"{o}dinger equation) in
$t$-dependent single- vs double-slit experiments recently discussed by the
group of Aharonov\cite{Tollaksen} who use a completely different method (with
modular variables in the Heisenberg picture). One should also note other
recent works such as \cite{He}, that rightfully emphasize that Physics cannot
currently predict how we dynamically go from the single-slit diffraction
pattern to the double-slit diffraction pattern (whether it is in a gradual and
causal manner or not) and where they propose relevant experiments to decide on
(measure) exactly this. Working with our nonlocal terms in such questions in
analogous experiments (i.e. by introducing (finite) scalar potentials on one
slit in a $t-$dependent way), in order to address the associated causal
issues, is currently under way.

\bigskip

\bigskip

\bigskip Students Kyriakos Kyriakou and Georgios Konstantinou of the
University of Cyprus and Areg Ghazaryan of Yerevan State University are
acknowledged for having carefully reproduced all results. Dr. Kleopatra
Christoforou of the Department of Mathematics and Statistics of the University
of Cyprus is acknowledged for a discussion concerning the mathematical method followed.

\textbf{Figure 1.} (Color online) Examples of the simplest
field-configurations (in simple-connected spacetimes), where the nonlocal
terms are nontrivial:

(a) a striped case in 1+1 spacetime, where the electric flux enclosed in the
\textquotedblleft observation rectangle\textquotedblright\ is dependent on t
but independent of x$\boldsymbol{;}$ (b) a triangular distribution in 2-D
space, where the part of the magnetic flux inside the \textquotedblleft
observation rectangle\textquotedblright\ depends on both x and y. The
appropriate choices for the corresponding nonlocal functions $g(x)$ and $h(y)$
are given in the text (eqns (\ref{triangular1}) and (\ref{triangular2})).


\bigskip

\bigskip

\bigskip

\bigskip

\begin{thebibliography}{99}                                                                                               %


\bibitem {Weyl}\bigskip H. Weyl, \textit{Z. Phys}. \textbf{56}, 330 (1929);
for a historical review see \ L. O'Raifeartaigh \& N. Straumann, \textit{Rev.
Mod. Phys}. \textbf{72}, 1 (2000)

\bibitem {AB}Y. Aharonov and D. Bohm, \textit{Phys. Rev}. \textbf{115}, 485 (1959)

\bibitem {Peshkin}M. Peshkin \& A. Tonomura, Lecture Notes in Physics 340,
Springer-Verlag (1989)$\boldsymbol{,}$ Part One

\bibitem {Tollaksen}J. Tollaksen, Y. Aharonov, A. Casher, T. Kaufherr, and S.
Nussimov, \ \ \ arXiv$\boldsymbol{:}$09010.4227 \ (Oct. 2009)

\bibitem {He}G. P. He, \ \ \ \ \ arXiv$\boldsymbol{:}$0907.1974 \ (Nov. 2009)

\bibitem {WernerBrill}F. G. Werner \& D. R. Brill, \ \textit{Phys. Rev. Lett}.
\textbf{4}, 344 (1960)

\bibitem {vanKampen}N. G. van Kampen, \textit{Phys. Lett.} \textbf{106A}, 5 (1984)

\bibitem {BrownHolland}H. R. Brown \& P. R. Holland, \textit{Amer. Journ.
Phys.} \textbf{67}, 204 (1999)

\bibitem {WuYang}T. T. Wu and C. N. Yang, \textit{Phys. Rev}. \textit{D}
\textbf{12}, 3845 (1975)

\bibitem {Iddings}P. D. Noerdlinger, \textit{Il Nuovo Cimento} \textbf{23},
158 (1962)

\bibitem {BrownHome}R.\ A. Brown \& D. Home, \textit{Il Nuovo Cimento}
\textbf{107B}, 303 (1992)

\bibitem {Troudet}T. Troudet, \textit{Phys. Lett.} \textbf{111A}, 274 (1985)

\bibitem {kyriakos}This has been first noted by graduate student K. Kyriakou

\bibitem {Feynman}R. P. Feynman, R. B. Leighton and M. Sands, \textit{The
Feynman Lectures on Physics}, vol. II, Chapter 15

\bibitem {Felsager}B. Felsager, \textquotedblleft\textit{Geometry, Particles,
and Fields}\textquotedblright, \ Springer-Verlag (1998), \ p.55

\bibitem {Batelaan}H. Batelaan \& A. Tonomura, \textit{Physics Today}
\textbf{62} (issue no. 9), p. 38 (September 2009)

\bibitem {LuanTang}P. G. Luan \& C. S. Tang, \ \textit{J. Phys.: Condens.
Matt. }\textbf{19}, 176224 (2007)

\bibitem {Compendium}\textit{Compendium of Quantum Physics}, Ed. D.
Greenberger \textit{et al}., Springer-Verlag 2009

\bibitem {Erlichson}H. Erlichson, \textit{Amer. Journ. Phys.} \textbf{38}, 162 (1970)

\bibitem {Silverman}I.e. M. P. Silverman, \textquotedblleft\textit{More than
One Mystery}\textquotedblright, Springer-Verlag (1995), p. 10$\boldsymbol{;}$
or, better, the same author's more recent \ \textquotedblleft\textit{Quantum
Superposition}\textquotedblright, Springer-Verlag (2008), \ p.13

\bibitem {Feynman2}Ref. [14], p. 15-13

\bibitem {Silverman2}First book of Ref. [20], p. 16$\boldsymbol{;}$ or second
book of Ref. [20], p. 19

%

\begin{align*}
& \\
& \\
& \\
& \\
& \\
& \\
& \\
& \\
& \\
& \\
&
\end{align*}%
\[
\]
%

\begin{align*}
& \\
&
\end{align*}


\ \ \ \ \ \ \ \ \ \ \ \ \ \ \ \ \ \ \ \ \ \ \ \ \ \ \ \ \ \ \ \ \ \ \ \ \ \ \ \ \ \textbf{FIGURE\ CAPTION}%

\end{thebibliography}
\end{document}